# An On/Off Berry Phase Switch in Circular Graphene Resonators

Fereshte Ghahari[1,2][*], Daniel Walkup[1,2][*], Christopher Gutiérrez[1,2][*], Joaquin F. Rodriguez-Nieva[3,4][*], Yue Zhao[1,2,5], Jonathan Wyrick[1], Fabian D. Natterer[1,6], William G. Cullen[1], Kenji Watanabe[7], Takashi Taniguchi[7], Leonid S. Levitov[3], Nikolai B. Zhitenev[1], and Joseph A. Stroscio[1][†]

[1]Center for Nanoscale Science and Technology, National Institute of Standards and Technology, Gaithersburg, MD 20899, USA
[2]Maryland NanoCenter, University of Maryland, College Park, MD 20742, USA
[3]Department of Physics, Massachusetts Institute of Technology, Cambridge, MA 02139, USA
[4]Department of Physics, Harvard University, Cambridge, MA 02138, USA
[5]Department of Physics, South University of Science and Technology of China, Shenzhen, China
[6] Institute of Physics, École Polytechnique Fédérale de Lausanne, Switzerland
[7]National Institute for Materials Science, Tsukuba, Ibaraki 305-0044, Japan

Abstract: The phase of a quantum state may not return to its original value after the system's parameters cycle around a closed path; instead, the wavefunction may acquire a measurable phase difference called the Berry phase. Berry phases typically have been accessed through interference experiments. Here, we demonstrate an unusual Berry-phase-induced spectroscopic feature: a sudden and large increase in the energy of angular-momentum states in circular graphene *p-n* junction resonators when a small critical magnetic field is reached. This behavior results from turning on a π-Berry phase associated with the topological properties of Dirac fermions in graphene. The Berry phase can be switched on and off with small magnetic field changes on the order of 10 mT, potentially enabling a variety of optoelectronic graphene device applications.

---

[*] These authors contributed equally to this work.
[†] To whom correspondence should be addressed.

Geometric phases are a consequence of a phenomenon that can be described as "global change without local change". A well-known classical example is the parallel transport of a vector around a path on a curved surface, which results in the vector pointing in a different direction after returning to its origin (Fig. 1A). Extending the classical phenomenon to the quantum realm by replacing the classical transport of a vector by the transport of a quantum state gave birth to the Berry phase, which is the geometric phase accumulated as a state evolves adiabatically around a cycle according to Schrödinger's equation (*1–5*). Since its discovery (*1*), non-trivial Berry phases have been observed in many quantum systems with internal degrees of freedom, such as neutrons (*6*), nuclear spins (*7*), and photons (*8*, *9*), through a Berry-phase-induced change in quantum interference patterns. Quantum systems in which a Berry phase alters the energy spectrum are comparatively rare; one example is graphene, where massless Dirac electrons carry a pseudospin ½, which is locked to the momentum **Π**. Because of the spin-momentum locking, the Berry phase associated with the state |**Π**⟩ can take only two values, 0 or $\pi$, which gives rise to the unconventional 'half-integer' Landau level structure in the quantum Hall regime (*10–12*). Yet, in most cases studied to date, the Berry phase in graphene has played a "static" role because controlling the trajectories (and hence |**Π**⟩) of graphene electrons is experimentally challenging. Recently introduced graphene electron resonators (*13–15*) enable exquisite control of the electron orbits by means of local gate potentials, and offer a unique opportunity to alter and directly measure the Berry phase of electron orbital states.

Here we report the control of the Berry phase of Dirac particles confined in a graphene electron resonator using a weak magnetic field, as recently suggested by theory (*16*). In this approach, a magnetic field enables fine control of the evolution of |**Π**⟩ around the Dirac point for individual resonator states (Fig.1B). A variation in the Berry phase,

which is accumulated during the orbital motion of the confined states, can be detected from changes in the energy dispersion of electron resonances. For this reason, we use scanning tunneling spectroscopy (STS) to directly measure the resonator-confined electronic states, giving direct access to the shifts in the quantum phase of the electronic states.

Graphene resonators confine Dirac quasiparticles by Klein scattering from *p-n* junction boundaries (*13–25*). Circular graphene resonators comprised of *p-n* junction rings host the whispering gallery modes analogous to those of acoustic and optical classical fields (*13–15*). These circular resonators can be produced by a scanning tunneling microscope (STM) probe in two different ways: (i) by using the electric field between the STM probe tip and graphene, with the tip acting as a moveable top gate (*13*), or (ii) by creating a fixed charge distribution in the substrate, generated by strong electric field pulses applied between the tip and graphene/boron nitride heterostructure (*14*). Both methods are used in our study of Berry-phase switching of resonator states.

Measurements were performed in a custom built cryogenic STM system (*26*) on two fabricated graphene/hBN/SiO$_2$ heterostructure devices, designed specifically for STM measurements (*27*). The field-induced switching of the graphene resonator states was initially observed in the movable tip-induced *p-n* junction resonators defined by the STM tip gating potential (*13*)(Section II of (*27*)). Subsequently, we created fixed *p-n* junction resonators, allowing us to study the spectroscopic properties and field dependence in a more controlled manner without a varying *p-n* junction potential; this report focuses mainly on the data obtained in this latter way. The fixed resonators display a pattern of Berry phase switching of the resonator modes that agrees with the one seen in the movable *p-n* junctions, demonstrating that different methods for creating the circular graphene resonators in separate devices lead to similar results.

Fixed circular-shaped *p-n* junctions were created by applying an electric field pulse between the STM probe and graphene device to ionize impurities in the hBN insulator, following Ref. (*28*). This method creates a stationary screening charge distribution in the hBN insulating under-layer, resulting in a fixed circular doping profile in the graphene sheet (Figs. 2, A and B) [see fig. S10 for schematic of the method]. We probe the quantum states in the graphene resonator by measuring the tunneling differential conductance, $g(V_b, V_g, r, B) = dI/dV_b$, as a function of tunneling bias, $V_b$, back gate potential, $V_g$, spatial position, $r$, and magnetic field, $B$. The quasi-bound resonances, originating from Klein scattering at the *p-n* ring, are seen in measurements made at $B = 0$ as a function of radial distance $r$ (Fig. 2D). The eigenstates in Fig. 2D form a series of resonant levels vertically distributed in energy, and are seen to exist within an envelope region of the confining *p-n* junction potential outlined by a high intensity state following the *p-n* junction profile. Because of rotational symmetry, the corresponding quantum states are described by radial quantum numbers, *n*, and angular momentum quantum numbers, *m* [Section I of (27)]. Only degenerate states with $m = \pm 1/2$ have non-zero wavefunction amplitude in the center of the resonator, as shown by the calculated wavefunctions in Fig. 2C, and dominate the measurements at $r = 0$ in Fig. 2D (*13–15*). Higher angular momentum states are observed off-center in the spatial distribution of the resonator eigenstates in Fig. 2D, but can be difficult to distinguish because of overlapping degenerate levels (*14*, *15*). Several calculated wavefunctions and corresponding eigenstates are indicated by color-coded circles in Figs. 2, C and D, respectively, to illustrate some patterns of the states in the spectroscopic map.

States with a specific radial quantum number $n$ follow arcs trailing the parabolic outline of the confining potential in Fig. 2D [see fig. S1 for the enumeration of the various $(n, m)$ quantum states].

We probe the magnetic field dependence of the $m = \pm 1/2$ resonator states (Fig. 3A) for the states $n = 1$ to $n = 5$ corresponding to the energy range indicated by the yellow dashed line in Fig. 2D. We see the degenerate $m = \pm 1/2$ states in the center of the map at $\boldsymbol{B} = 0$, corresponding to the states seen in the center of Fig. 2D at $\boldsymbol{r} = 0$. However, as the magnetic field is increased, new resonances suddenly appear between the $n$th quantum levels, where none previously existed. The spacing between the new magnetic field induced states, $\delta\varepsilon_m$, is about one half the spacing, $\Delta\varepsilon$, at $\boldsymbol{B} = 0$. A magnified view of the map around the $n = 4$ level (Fig. 3C) shows the appearance of new resonances switching on at a critical magnetic field, $B_C \approx \pm 0.11$ T. As explained below, the switching transition corresponds to the sudden separation of the $m = \pm 1/2$ sublevels, which is very sharp [see Fig. 3D, and fig. S11 for more detail]. This energy separation, on the order of 10 meV at $B = 0.1$ T, is much larger than other magnetic field splittings, i.e. Zeeman splitting ($\varepsilon_Z = \mu_B B_C \approx 0.01$ meV, with $\mu_B$ the Bohr magneton) or orbital effects [$\varepsilon_{\text{orb}} \approx 1$ meV, see Ref. (*16*)]. As we discuss below, these results correspond to the energy shift of particular quantum states, which suddenly occurs owing to the switching on of a $\pi$ Berry phase when a weak critical magnetic field is reached.

A simple understanding of the sudden jump in energy observed in the experiment (Fig. 3A) can be drawn from the Bohr-Sommerfeld quantization rule, which determines the energy levels $E_n$:

$$\frac{1}{\hbar}\oint_{C_R(E_n)} \boldsymbol{p} \cdot d\boldsymbol{q} = 2\pi\left(n + \gamma - \frac{\varphi_B}{2\pi}\right). \tag{1}$$

Here $\boldsymbol{q}$ and $\boldsymbol{p}$ are the canonical coordinates and momenta, respectively, $\hbar$ is Planck's constant divided by $2\pi$, $\varphi_B$ is the Berry phase accumulated in each orbital cycle, $C_R$ is a phase-space contour described below, and $\gamma$ is a constant, the so-called Maslov index (5). The orbits are obtained from the semiclassical Hamiltonian $H = E_n = v_F|\boldsymbol{\Pi}| + U(r)$, with $U(r)$ the confining potential and $\boldsymbol{\Pi} = \boldsymbol{p} - e\boldsymbol{A}$ the kinetic momentum ($e$ is the electron charge and $\boldsymbol{A}$ the vector potential). At zero magnetic field, states with opposite angular momenta $\pm m$ are degenerate, and their orbits are time-reversed images of one another (Fig.1C, left). When a small magnetic field is turned on, the Lorentz force bends the paths of the $+m$ and $-m$ charge carriers in opposite directions, breaking the time-reversal symmetry and slightly lifting the orbital degeneracy. Crucially, at a critical magnetic field $B_c$, the orbit with angular momentum anti-parallel to the field is twisted by the Lorentz force into a qualitatively different "skipping" orbit (Fig. 1C, right). At this transition, the momentum-space trajectory, defined below, encloses the Dirac point, and $\varphi_B$ discontinuously jumps from $0$ to $\pi$; states with $\pm m$, which are degenerate at $B = 0$, are abruptly pulled apart by half a period.

The intuitive picture described above can be made rigorous by using the Einstein-Brillouin-Keller (EBK) quantization (29, 30). Besides being more rigorous, EBK quantization facilitates visualization of the trajectory of $\boldsymbol{\Pi}$ along a semiclassical orbit, particularly because the orbits are quasiperiodic. In central force motion, the particle's orbit takes place in an annulus between the two classical turning points, and because the momentum in this annulus is two-valued, one can define a torus on which the momentum is uniquely determined [Section I of (27)]. The EBK quantization rules are formed by evaluating $\oint \mathbf{p} \cdot d\mathbf{q}$ along the two topologically distinct loops on this torus, $C_R$ and $C_\theta$ (Fig. 1D). The EBK rule along $C_\theta$ gives the half-integer quantization of angular momentum; that along $C_R$ (Eq. 1) determines the energy levels for a given angular momentum. The Berry phase term in Eq. 1 is determined by the winding number of $\mathbf{\Pi}$ about the origin, evaluated on $C_R$. Below $B_c$, the azimuthal component of $\mathbf{\Pi}$ has the same sign along $C_R$ (Fig. 1E, left): the corresponding $\mathbf{\Pi}$-space loop (Fig. 1B, blue curve) does not enclose the origin and the Berry phase is zero. Above $B_c$, however, $\mathbf{\Pi}$ has a sign change along $C_R$ (Fig. 1E, right): the loop encircles the Dirac point and the Berry phase is $\pi$ (Fig. 1B, red curve). Fig. 1B provides an intuitive visualization of the switching mechanism: the changing magnetic field shifts the $\mathbf{\Pi}$-space contour, and at $B_c$ it slips it over the Dirac cone apex, instantly changing the right side of Eq. 1 by $\pi$ and shifting the energy levels accordingly [see Movie S1].

The semiclassical picture additionally allows us to estimate the strength of the critical field $B_c$ necessary to switch the Berry phase (16). The value of $B_c$, which is sensitive to the confining potential profile, is obtained by finding the field strength $B$ necessary to bend the electron orbit into a skipping orbit at the outer return point $r_o$, thus resulting in zero azimuthal momentum: $\Pi_\theta = m\hbar/r_o - eB_c r_o/2 = 0$. For a parabolic potential $U(r) = \kappa r^2$, which accurately describes the low energy resonances (16), one obtains a simple expression:

$$B_c = \frac{2\hbar}{e}\frac{m\kappa}{\varepsilon}, \qquad (2)$$

where $\varepsilon$ is the energy of the orbit. Using $\kappa = 10$ eV/μm² obtained from a parabolic fitting of the potential profile (see Fig.2D) and $\varepsilon = 65$ meV (relative to the position of the Dirac point), which corresponds to the $(m = 1/2, n = 1)$ state in Fig.3A, we find $B_c = 0.1$ T, in excellent agreement with the magnetic field values measured in the experiment.

A detailed comparison between experiment and theory can be made by solving the 2D Dirac Equation describing graphene electrons in the presence of a confining potential $U(r)$ and a magnetic field $B$, $[v_F(\boldsymbol{\sigma} \cdot \boldsymbol{\Pi}) + U(r)]\Psi(r) = \varepsilon\Psi(r)$ Following the semiclassical discussion above, we use a simple parabolic potential model $U(r) = \kappa r^2$, with $\kappa = 10$ eV/µm² to fit the experimental data. The calculated local density of states (LDOS) at the center of the resonator [see Section I of (27) for details (27)] is in good agreement with the experiment (Figs. 3, A and B) and exhibits the half-period jumps between time-reversed states at similar magnetic field values.

Equation 2 also predicts a higher critical field for larger *m* states, as a larger Lorentz force is needed to induce skipping orbits. These larger angular momentum states have zero wavefunction weight at $r = 0$ (see Fig.2C), but can be probed at positions away from the center of the resonator (*14*, *15*). Figure 4A shows the spectral conductance map measured at a position of 70 nm away from the center of the *p-n* junction resonator [see fig. S12 for additional measurements off center], and Fig. 4B shows the calculated LDOS at the same position. Much more complex spectral features are seen in comparison with the on-center measurement (Fig. 3A), which is sensitive only to the $m = \pm 1/2$ states. First, at $B = 0$, twice as many resonances are observed within the same energy range when compared to Fig. 3A. This confirms that additional *m* states are contributing to the STM maps. Second, both the theory and experimental maps exhibit a "stair case" pattern. Such an arrangement results from an overlap between the field-jumped *m* state and the next higher non-jumped *m* state, which are nearly degenerate in energy. For example, at positive fields, the field jumped $m = 1/2$ state overlaps in energy with the non-jumped $m = 3/2$ state, giving rise to a strong intensity at this energy. Figures 4, C and D illustrate this effect by showing separately the $m = 1/2$ and $m = 3/2$ contributions to the total density of states, respectively; $B_c$ is indicated with a dashed line. In a similar fashion, this behavior continues for all adjacent *m* levels resulting in the series of staircase steps in the measured and calculated conductance maps in Fig. 4, A and B. Since the states jump by the half of the energy spacing, the staircase patterns can visually form upward and downward looking lines depending on subtle intensity variations at the transitions. For instance, while Fig. 4A shows mostly downward staircase patterns, other measurements (fig. S12) show both down and upward connections at the transitions; the parabolic potential model, Fig. 4B, shows an upward staircase. These behaviors depend on subtle effects, such as the potential shape in the off-center measurement, which are not relevant for our main discussion. More importantly, we plot, as a guide to the eye, $B_c$ in Eq. 2 (dashed lines) for different values of $m$, showing excellent agreement with the semiclassical estimation above (note that upon summation in Fig. 4B, the positions of the steps, i.e. the white fringes, seem shifted to higher *B* values; this is expected and analogous to the peak shift observed when summing two Lorentzian functions). Overall, our one-parameter Dirac equation gives a very good description of the resonance dispersion, the critical field $B_c$, and its dependence on energy and momentum.

Implications of these results include possible use in helicity-sensitive electro-optical measurements at THz frequencies with the ability to switch circular-polarized optical signals with small modulations of magnetic fields on the order of 10 mT (Section III.C of (*27*)). These applications, in conjunction with the fidelity of fabricating a variety of *p-n* junction and other electrostatic boundaries with impurity doping of boron nitride in graphene heterostructures, will expand the quantum tool box of graphene-based electron optics for future studies and applications.

**FIGURE CAPTIONS**

**Fig. 1. Dynamics of whispering gallery modes in circular graphene resonators**. (**A**) The parallel transport of a vector around a closed path *C* on a curved surface. For parallel transport the vector *t* remains perpendicular to *r*, and the orthogonal frame containing *t* and *r* does not twist about *r* (*3*). The transport results in the angular difference α between initial and final *t* vectors. (**B**) (lower) Schematic momentum-space contours for magnetic fields below (blue) and above (red) the critical magnetic field $B_c$, corresponding to the vectors shown in (D); (upper) the same contours projected on the Dirac cone by evaluating the kinetic energy $T(\Pi)$. Above $B_c$ (red), the contour encloses the Dirac point leading to a π Berry phase. (**C**) Schematics of the potential profile in the circular graphene resonator formed by a *p*-doped graphene center region and *n*-doped background, and classical orbits for positive *m* states below (left, blue) and above (right, red) $B_c$. (**D**) Schematic phase-space tori corresponding to the orbits shown in (C). The kinematic momenta $\Pi$ (arrows), uniquely defined on the torus, are shown along the topologically distinct loops $C_\theta$ and $C_R$. Lower panel: For $C_R$ below $B_c$ (blue), the winding number of $\Pi$ is zero and has a zero Berry phase; above $B_c$ (red), the winding number is one, leading to a π Berry phase.

**Fig. 2. Quantum whispering gallery modes of a graphene circular resonator.** (**A**) Schematic of the potential profile formed by (**B**) a *p*-doped graphene center region and *n*-doped background, created by ionizing impurities in the underlying hBN insulator. (**C**) Calculated

wavefunction components of a circular graphene resonator for a parabolic potential for various indicated $(n,m)$ states. (All scale bars 100 nm). (**D**) Differential conductance map vs. radial spatial position obtained from an angular average of an *xy* grid of spectra obtained over the graphene resonator. $m = \pm 1/2$ states appear in the center at $r = 0$, whereas states with higher angular momentum occupy positions away from center in arcs of increasing *m* values and common *n* value, as seen by the associated wavefunctions in (C) [see also fig. S1]. The solid blue line shows a parabolic potential with $\kappa$=10 eV/μm² and a Dirac point of 137 mV, which is used as a confining potential in the simulations shown in Figs. 3B and Figs. 4,B-D. The dashed yellow line at *r*=0, and dashed green line at *r*=70 nm, indicate the measurement positions for Figs. 3A and 4A.

**Fig. 3. The On/Off Berry phase switching in graphene circular resonators.** (**A**) Differential tunneling conductance map vs. magnetic field measured in the center of the graphene resonator for the *n*=1 to *n*=5 modes, indicated by the yellow dashed line in Fig. 2D. New resonances suddenly appear at a critical magnetic field of $B_C \approx \pm 0.11$ T in between those present at $B = 0$ T. (**B**) Calculated local density of states (LDOS) of the graphene resonator states in a magnetic field to compare with experiment in (A) using a parabolic confining potential with $\kappa = 10$ eV/μm² and Dirac point $E_D = 137$ meV (Fig. 2D). (**C**) Magnified view of the *n* = 4 resonance from (A) showing the Berry-phase induced jumping of the $m = \pm 1/2$ modes with magnetic field: $m = +1/2$ only jumps at positive fields, and $m = -1/2$ only at negative fields. The 2D maps in A-C are shown in the 2$^\text{nd}$ derivative to remove the graphene dispersive background. (**D**) The difference in energy between the $m = \pm 1/2$ states, $\delta\varepsilon_m$, for the *n* = 4 resonance vs. magnetic field. The right axis is in units of the energy difference between the *n*=4 and *n*=3 resonances, $\Delta\varepsilon$, at *B* = 0 T. The uncertainty reported represents one standard deviation

and is determined by propagating the uncertainty resulting from a least square fit of Lorentzian functions used to determine the peak position of the resonator modes. The Dirac cones schematically illustrate the switching action: At low fields, the switch is open with zero Berry phase. When the critical field is reached, a π Berry phase is turned on closing the switch.

**Fig. 4. Berry phase switching of higher angular momentum states**. (**A**) Differential tunneling conductance map vs. magnetic field measured 70 nm off-center of the graphene resonator. Kinks observed in the conductance correspond to the higher *m* modes which are seen to switch with higher critical magnetic fields. The staircase switching pattern results from *m* states that jump and overlap in energy the next higher non-switched *m* state, leading to an increased intensity at that energy (see text). Calculated LDOS vs magnetic field for a position 70 nm off center in a graphene circular resonator for (**B**) all *m* states, (**C**) $m = +1/2$, and (**D**) $m = +3/2$. The fan of dashed lines are calculations for the critical magnetic field using Eq. 2 with $\kappa = 10$ eV/μm$^2$ and $E_D = 137$ meV and *m* values as: (A),(B) $m = \pm\frac{1}{2}$ to $m = \pm 7/2$, and single *m* values indicated on top of (C,D). The maps are shown in the 2$^{nd}$ derivative (arbitrary values). See fig. S12 for additional off-center measurements.

**REFERENCES AND NOTES**

1. M. V. Berry, Quantal Phase Factors Accompanying Adiabatic Changes. *Proc. R. Soc. Lond. Math. Phys. Eng. Sci.* **392**, 45–57 (1984).

2. M. Berry, Anticipations of the Geometric Phase. *Phys. Today*. **43**, 34–40 (1990).

3. F. Wilczek, A. Shapere, *Geometric Phases in Physics* (WORLD SCIENTIFIC, 1989; http://www.worldscientific.com/worldscibooks/10.1142/0613), vol. 5 of *Advanced Series in Mathematical Physics*.

4. J W Zwanziger, M Koenig, and A. Pines, Berry's Phase. *Annu. Rev. Phys. Chem.* **41**, 601–646 (1990).

5. D. Xiao, M.-C. Chang, Q. Niu, Berry phase effects on electronic properties. *Rev. Mod. Phys.* **82**, 1959–2007 (2010).

6. T. Bitter, D. Dubbers, Manifestation of Berry's topological phase in neutron spin rotation. *Phys. Rev. Lett.* **59**, 251–254 (1987).

7. R. Tycko, Adiabatic Rotational Splittings and Berry's Phase in Nuclear Quadrupole Resonance. *Phys. Rev. Lett.* **58**, 2281–2284 (1987).

8. R. Y. Chiao, Y.-S. Wu, Manifestations of Berry's Topological Phase for the Photon. *Phys. Rev. Lett.* **57**, 933–936 (1986).

9. A. Tomita, R. Y. Chiao, Observation of Berry's Topological Phase by Use of an Optical Fiber. *Phys. Rev. Lett.* **57**, 937–940 (1986).

10. K. S. Novoselov *et al.*, Two-dimensional gas of massless Dirac fermions in graphene. *Nature*. **438**, 197–200 (2005).

11. Y. Zhang, Y.-W. Tan, H. L. Stormer, P. Kim, Experimental observation of the quantum Hall effect and Berry's phase in graphene. *Nature*. **438**, 201–204 (2005).

12. D. L. Miller *et al.*, Observing the quantization of zero mass carriers in graphene. *Science*. **324**, 924–927 (2009).

13. Y. Zhao *et al.*, Creating and probing electron whispering-gallery modes in graphene. *Science*. **348**, 672–675 (2015).

14. J. Lee *et al.*, Imaging electrostatically confined Dirac fermions in graphene quantum dots. *Nat. Phys.* **12**, 1032–1036 (2016).

15. C. Gutiérrez, L. Brown, C.-J. Kim, J. Park, A. N. Pasupathy, Klein tunnelling and electron trapping in nanometre-scale graphene quantum dots. *Nat. Phys.* **12**, 1069–1075 (2016).

16. J. F. Rodriguez-Nieva, L. S. Levitov, Berry phase jumps and giant nonreciprocity in Dirac quantum dots. *Phys. Rev. B*. **94**, 235406 (2016).

17. A. V. Shytov, M. S. Rudner, L. S. Levitov, Klein Backscattering and Fabry-Pérot Interference in Graphene Heterojunctions. *Phys. Rev. Lett.* **101**, 156804 (2008).

18. A. F. Young, P. Kim, Quantum interference and Klein tunnelling in graphene heterojunctions. *Nat. Phys.* **5**, 222–226 (2009).

19. V. V. Cheianov, V. Fal'ko, B. L. Altshuler, The Focusing of Electron Flow and a Veselago Lens in Graphene p-n Junctions. *Science*. **315**, 1252–1255 (2007).

20. A. De Martino, L. Dell'Anna, R. Egger, Magnetic Confinement of Massless Dirac Fermions in Graphene. *Phys. Rev. Lett.* **98**, 066802 (2007).

21. P. E. Allain, J. N. Fuchs, Klein tunneling in graphene: optics with massless electrons. *Eur. Phys. J. B*. **83**, 301–317 (2011).

22. L. C. Campos *et al.*, Quantum and classical confinement of resonant states in a trilayer graphene Fabry-Pérot interferometer. *Nat. Commun.* **3**, 1239 (2012).

23. A. Varlet *et al.*, Fabry-Pérot Interference in Gapped Bilayer Graphene with Broken Anti-Klein Tunneling. *Phys. Rev. Lett.* **113**, 116601 (2014).

24. J.-S. Wu, M. M. Fogler, Scattering of two-dimensional massless Dirac electrons by a circular potential barrier. *Phys. Rev. B*. **90**, 235402 (2014).

25. N. A. (Garg), S. Ghosh, M. Sharma, Scattering of massless Dirac fermions in circular p-n junctions with and without magnetic field. *J. Phys. Condens. Matter*. **26**, 155301 (2014).

26. R. J. Celotta *et al.*, Invited Article: Autonomous assembly of atomically perfect nanostructures using a scanning tunneling microscope. *Rev. Sci. Instrum.* **85**, 121301 (2014).

27. Additional supplementary text and data are available on Science Online.

28. J. Velasco *et al.*, Nanoscale Control of Rewriteable Doping Patterns in Pristine Graphene/Boron Nitride Heterostructures. *Nano Lett.* **16**, 1620–1625 (2016).

29. A. Einstein, On the Quantum Theorem of Sommerfeld and Epstein. *Deutshe Phys. Ges.* **19**, 82–92 (1917).

30. A. D. Stone, Einstein's unknown insight and the problem of quantizing chaos. *Phys. Today*. **58**, 37–43 (2005).

31. C. R. Dean *et al.*, Boron nitride substrates for high-quality graphene electronics. *Nat Nano*. **5**, 722–726 (2010).

32. A. C. Ferrari, Raman spectrum of graphene and graphene layers. *Phys Rev Lett*. **97**, 187401 (2006).

**ACKNOWLEDGMENTS**

F.G., D.W., C.G., and Y. Z. acknowledges support under the Cooperative Research Agreement between the University of Maryland and the National Institute of Standards and Technology Center for Nanoscale Science and Technology, Grant No. 70NANB10H193, through the University of Maryland. J.W. acknowledges support from the Nation Research Council Fellowship. F. D. N. greatly appreciates support from the Swiss National Science Foundation under project number PZ00P2_167965. Y. Z. acknowledges support by the National Science Foundation of China under project number 11674150 and the National 1000 Young Talents Program. J.F.R.-N. acknowledges support from the NSF grant DMR-1507806. K.W. and T.T.

acknowledge support from JSPS KAKENHI grant no. JP15K21722. L.S.L. acknowledges support by the Center for Integrated Quantum Materials (CIQM) under NSF award 1231319 and by the Center for Excitonics, an Energy Frontier Research Center funded by the U.S. Department of Energy, Office of Science, Basic Energy Sciences, under award no. DESC0001088. We thank Steve Blankenship and Alan Band for their contributions to this project, and we thank Michael Zwolak and Mark Stiles for valuable discussions.

**SUPPLEMENTARY MATERIALS**

Supplementary Text
Figs. S1 to S12
References (31,32)
Movie S1

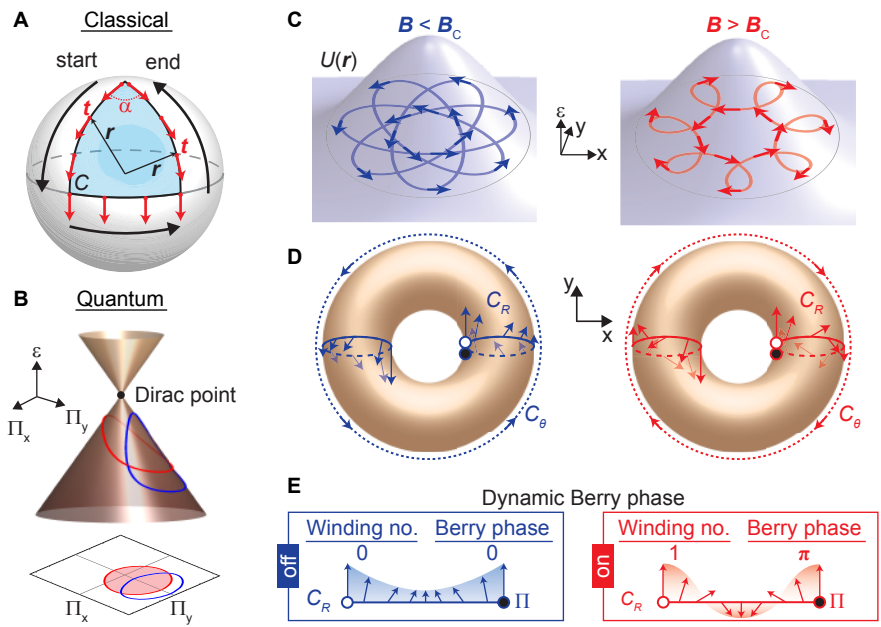

Figure 1

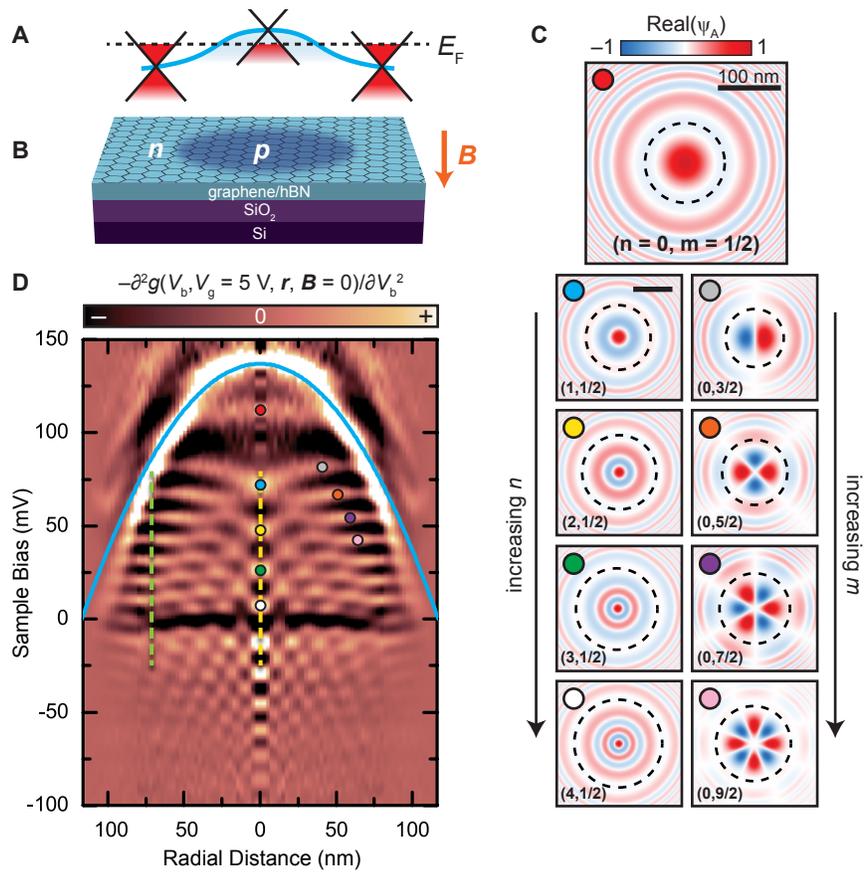

Figure 2

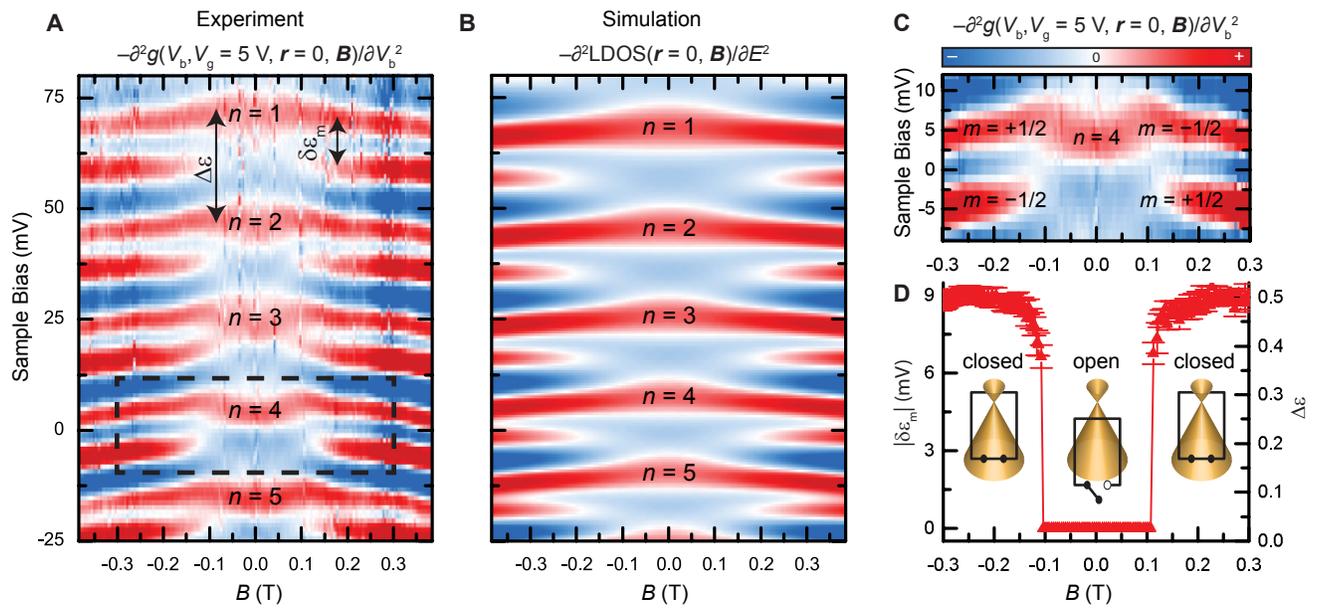

Figure 3

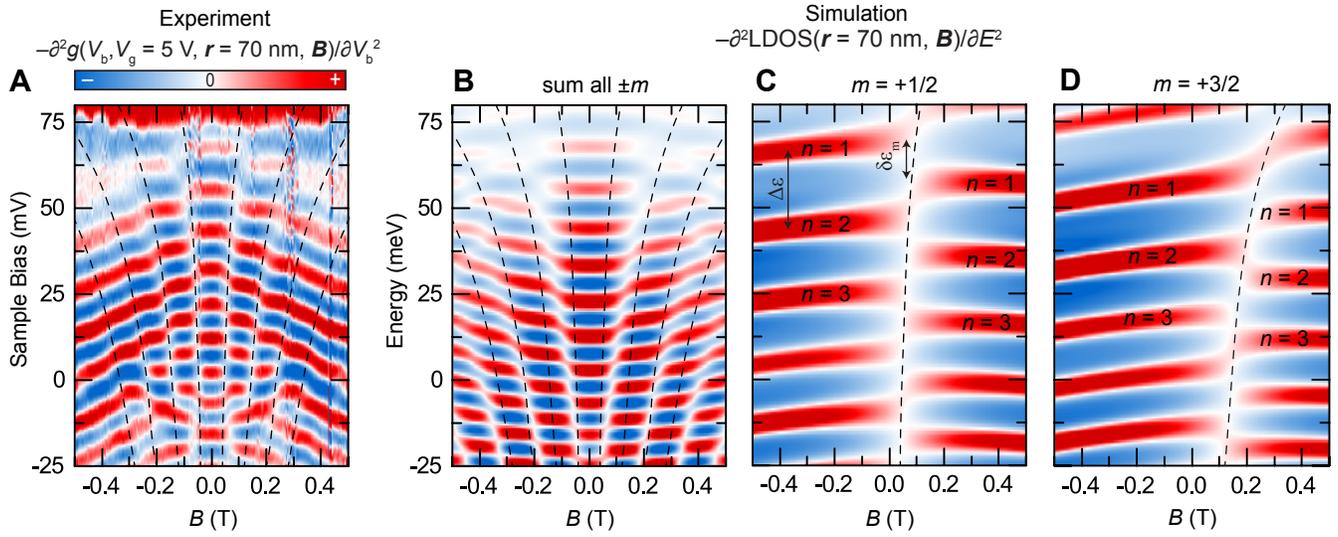

Figure 4

# Supplementary Materials for

## An On/Off Berry Phase Switch in Circular Graphene Resonators

Fereshte Ghahari[1,2*], Daniel Walkup[1,2*], Christopher Gutiérrez[1,2*], Joaquin F. Rodriguez-Nieva[3,4*], Yue Zhao[1, 2,5], Jonathan Wyrick[1], Fabian D. Natterer[1,6], William G. Cullen[1], Kenji Watanabe[7], Takashi Taniguchi[7], Leonid S. Levitov[3], Nikolai B. Zhitenev[1], and Joseph A. Stroscio[1]

[1]Center for Nanoscale Science and Technology, National Institute of Standards and Technology, Gaithersburg, MD 20899, USA
[2]Maryland NanoCenter, University of Maryland, College Park, MD 20742, USA
[3]Department of Physics, Massachusetts Institute of Technology, Cambridge, MA 02139, USA
[4]Department of Physics, Harvard University, Cambridge, MA 02138, USA
[5]Department of Physics, South University of Science and Technology of China, Shenzhen, China
[6] Institute of Condensed Matter Physics, École Polytechnique Fédérale de Lausanne, Switzerland
[7]National Institute for Materials Science, Tsukuba, Ibaraki 305-0044, Japan.

correspondence to:  joseph.stroscio@nist.gov

**This PDF file includes:**

- I. Theory of Circular Graphene Resonators
- II. Circular Graphene Resonators Made with Tip Gating
- III. Circular Graphene Resonators Made with Impurity Doping of the hBN Insulator
- IV. Full Reference List
  Figures S1 to S12
  Caption for Movie S1

**Other Supplementary Materials for this manuscript includes the following:**

Movie S1

---

[*] These authors contributed equally to this work.

## I. Theory of Circular Graphene Resonators

We summarize the calculations for circular graphene resonators based on both quantum and classical descriptions. More detailed descriptions can be found in references (*13, 16*).

### A. Quantum calculations of circular graphene resonators.

For a quantum description of the graphene resonator states, we use the effective graphene Hamiltonian,

$$\varepsilon \psi(r) = [v_F \boldsymbol{\sigma} \cdot \boldsymbol{\Pi} + U(r)]\psi(r), \tag{S1}$$

where $\Pi_{x,y} = -i\hbar \partial_{x,y} - eA_{x,y}$ is the kinematic momentum, $\boldsymbol{\sigma} = (\sigma_x, \sigma_y)$ are pseudospin Pauli matrices and $U(r)$ is the confining electrostatic potential. A simple parabolic potential model, $U(r) = \kappa r^2$, provides a good description of the potential profile for low energy resonant states, as can be seen by the outline of the quantum dot states shown in Fig. S1. The eigenstates of Eq. S1 can be expressed using polar decomposition, taking the axial gauge $A_x = -By/2, A_y = Bx/2$ to preserve rotational symmetry,

$$\psi_m(r, \theta) = \frac{e^{im\phi}}{\sqrt{r}} \begin{pmatrix} u_1(r)e^{-i\theta/2} \\ iu_2(r)e^{i\theta/2} \end{pmatrix}, \tag{S2}$$

with *m* a half integer. This decomposition allows one to rewrite Eq. (S1) as,

$$\begin{pmatrix} r^2 - \varepsilon & \partial_r + m/r - Br/2 \\ -\partial_r + m/r - Br/2 & r^2 - \varepsilon \end{pmatrix} \begin{pmatrix} u_1 \\ u_2 \end{pmatrix} = 0 \tag{S3}$$

where *r*, *B*, and $\varepsilon$ are in units of $r_*$, $B_*$, $\varepsilon_*$:

$$r_* = \sqrt[3]{\hbar v/\kappa} \approx 40 \text{ nm}, \; \varepsilon_* = \sqrt[3]{(\hbar v)^2 \kappa} \approx 16 \text{ meV}, \; B_* = (\hbar/e)\sqrt[3]{(\kappa/\hbar v)^2} \approx 0.4 \text{ T}. \tag{S4}$$

Here we used typical values of $v \approx 10^6$ m/s and $\kappa = 10$ eV/µm$^2$.

To compare with the experimental *dI/dV* measurements, we calculate the local density of states (LDOS) as a sum of *m*-state contributions $D(\varepsilon) = \sum_m D_m(\varepsilon)$, with

$$D_m(\varepsilon) = \sum_\alpha \langle |u_\alpha(r = r_0)|^2 \rangle_{\lambda_d} \delta(\varepsilon - \varepsilon_\alpha). \tag{S5}$$

Here α labels the radial eigenstates of Eq. S3 for fixed *m*. The factor $\langle |u_\alpha(r = r_0)|^2 \rangle_{\lambda_d} = \int_0^\infty dr' |u_\alpha(r')|^2 e^{-(r'-r_0)^2/2\lambda_d^2}$ provides a spatial average of the wave function with a Gaussian weight to account for the finite size of the tunneling region.

Figure S1 shows the measured differential conductance measurement from Fig. 2D, where we assign the various eigenstates. The eigenstates for the circular resonator form a system of levels associated with radial and angular momentum quantum numbers, *n,m*, distributed both vertically in energy and horizontally as a function of spatial position. The eigenstates are outlined by a strong scattering state which follows the potential dispersion, with the Dirac point close to 137 mV at the center of the quantum dot, indicated by the parabolic potential overlaid in the black lines. For each eigenstate $\psi_{n,m}$ there are *n*+1 maxima in radial wavefunction, however, most of the weight is on the first maximum (*14*). For fixed *n*, the main maxima in the eigenstates form bands of arcs following the shape of the potential profile, with *n*=0 closest to the outside envelope, as shown in Fig. S1A. With increasing *n*, the bands come closer to the center. For each *n* band, the progression of the states starts with *m*=1/2 at the center *r*=0, and increases in integer values to 3/2, 5/2, etc. In the center of the resonator, the states are all *m*=1/2, and with increasing *n* values. Each of these *n* values starts a progression, which gives rise to the bending outlined in Fig. S1A. Similarly, there are bands of states with common *m* value with increasing *n* values, starting at the top of the map, as shown in Fig. S2B.

When these two sets of arcs are combined, they enumerate the main maxima of (*n,m*) eigenstates and overlap the maxima observed in the conductance map in Fig. S1C.

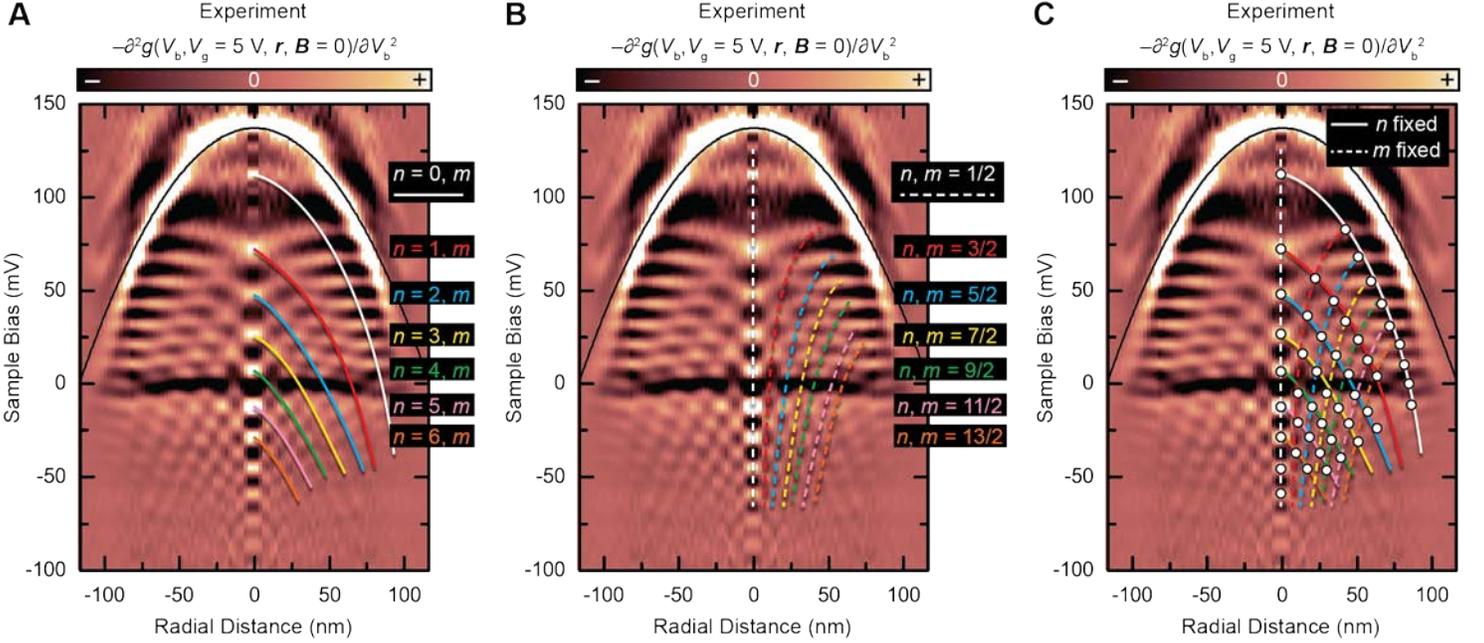

**Figure S1: Identifying the eigenstates of a graphene circular resonator.** The differential conductance measurement of the circular graphene resonator from Fig. 2D. The solid black line shows a quadratic potential with $\kappa=10$ eV/μm$^2$ and a Dirac point of 137 mV, which is used as a confining potential in the simulations. (**A**) The solid color lines indicate the family of states with a strong maximum, and common $n$ value and increasing $m$ values, originating at $m=1/2$ at $r=0$. These maxima make arcs which follow the outline of the confining potential. (**B**) The dashed color lines indicate the family of states with common $m$ values and increasing $n$ starting at $n=0$ at the top of the map. (**C**) Superimposing the two sets of colored lines from (A) and (B) show that the maxima observed in the conductance map are observed at the intersection of the colored lines and enumerate the main maxima in the ($n,m$) eigenstates for the resonator.

B. Classical mechanics simulations of the circular resonator

To model the semiclassical orbits (Fig. 1), we performed dynamics calculations using the Hamiltonian

$$H(r, \phi, p_r, p_\phi) = v_F \sqrt{p_r^2 + \left(\frac{p_\phi}{r} - \frac{eB}{2}r\right)^2} + \kappa r^2, \tag{S6}$$

in which $v_F$, $e$, $B$, and $\kappa$ are parameters. The distance is measured in Angstroms, time in seconds, energy in eV.

For the orbits in Fig. 1C of the main text, the particle was started with the initial conditions {$r$=200Å, $\phi$=0, $p_r$=0, $p_\phi$=$\hbar/2$} and run for < 1ps in magnetic fields specifically chosen to give (approximately) closed orbits: $B$=-7 mT on the left, and $B$=+514 mT on the right. The loops in Fig. 1B were derived from the momentum data of the same orbits as follows: we define the radial and azimuthal kinematic momentum

$$\Pi_r = p_r, \tag{S7}$$

$$\Pi_\phi = \frac{1}{r}\left(p_\phi - \frac{eBr^2}{2}\right); \tag{S8}$$

a plot of $\Pi_r$ vs. $\Pi_\phi$ gives the loops of Fig. 1B, lower. The z-coordinate in Fig. 1B, upper, is the (negative) kinetic energy $T = -v_F\sqrt{\Pi_r^2 + \Pi_\phi^2}$.

For the orbits in the movie, the initial conditions were chosen in a slightly different way: instead of using a fixed $r_0$, for each $B$ we calculate the energy necessary to give the orbit a fixed radial action $J_r$=$h/2$, and set $r_0$ equal to the inner turning point. This matches the quantization rule for $B < B_c$, so in that regime the orbit may be regarded as belonging to the $n=0$ quantum state. In the movie S1 we used $\kappa=10$ eV/µm²; in Fig. 1 we used $\kappa=4$ eV/µm².

### C. Einstein-Brillouin-Keller Quantization rules

As mentioned in the main text, in central force motion the momentum $\boldsymbol{p}(\boldsymbol{q})$ is a two-valued vector field, defined in the annulus between the inner and outer turning points (Fig. S2A). Given that $p_\phi$ is conserved, the field is completely specified from $p_r(r)$. For our Hamiltonian (S6)

$$p_r(r) = \pm \sqrt{\left(\frac{\varepsilon - \kappa r^2}{v_F}\right)^2 - \left(\frac{p_\phi}{r} - \frac{eB}{2}r\right)^2}. \tag{S9}$$

Fig. S2B shows a typical curve $p_r(r)$, calculated using $\kappa=10$ eV/µm², $p_\phi = \hbar/2$ (i.e., $m=1/2$), and $\varepsilon = 54$ meV. In the torus of the main text (Fig. 1D), the poloidal loop $C_R$ extends from the inner turning point to the outer and back again, so the corresponding EBK quantization rule (Eq. 1) becomes

$$\oint_{C_R} \boldsymbol{p} \cdot d\boldsymbol{q} = 2 \int_{r_{in}}^{r_{out}} |p_r(r)| dr = 2\pi\hbar \left(n + \gamma - \frac{\varphi_B}{2\pi}\right); \tag{S10}$$

i.e. it takes the usual form of a 1D semiclassical quantization rule.

A more explicit picture of the EBK torus can be obtained by regarding $p_r(r)$ as the toroidal cross-section. A particle governed by the Hamiltonian (S6) moves through a 3D phase space of $r$, $\phi$, $p_r$, and always remains on the toroidal surface so defined. If the orbit is not closed, it will eventually cover the entire surface. In this picture the EBK quantization rule (Eqs. 1 and S10) becomes a quantization rule on the cross-sectional area of the torus.

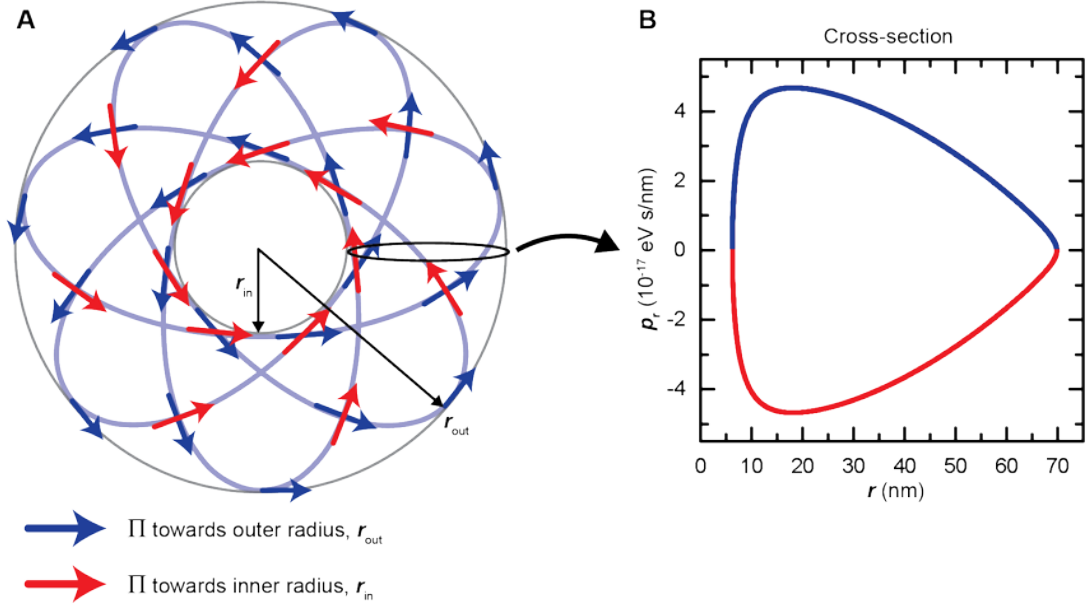

**Figure S2: Two-valued momentum field and toroidal cross-section**. (**A**) Top view of a calculated closed orbit, the same as in Fig. 1C, left. The inward and outward momentum vectors are shown schematically as red and blue arrows respectively. (**B**) A calculated $p_r(r)$ curve, with inward and outward motions colored red and blue respectively. The curve was calculated using $\kappa$=10 eV/µm², $p_\phi = \frac{\hbar}{2}$, and $\varepsilon = 54$ meV, corresponding to a radial action of $\oint p_r dr = \hbar\pi$.

## II. Circular Graphene Resonators Made with Tip Gating

The first method we used to make a circular *p-n* junction graphene resonator for STM measurements involved the field-effect gating from the probe tip potential to locally dope the graphene sheet relative to the background doping (*13*). This method results in a *p-n* junction graphene quantum dot which travels with the position of the STM probe tip. As a result, measurements can only be made in the center of the resonator. The *p-n* junction rings can confine whispering gallery modes only in certain regions of the phase space defined by the sample bias and back gate voltage, where a sufficient tip-graphene potential difference is needed, as described in Ref. (*13*). The Berry phase switching observed in the fixed graphene quantum dot in the main text was also seen in the traveling dots, albeit with the complexity that the confining potential is a dynamic function of the sample bias $V_b$ and back gate voltage $V_g$, which are both used to create and map the resonances, as shown in Fig. S3-S5. These measurements show that the effect of the tip potential on the fixed dot measurements is negligible, since potentials in excess of 300 mV are needed to establish the resonator states when the contact potential between probe and graphene is not large (*13*). Thus, the bias range of ±100 mV used in the fixed dot measurement does not create a tip-induced quantum dot in the graphene and the resonator states seen with the hBN impurity doping are largely unaffected by the low values of sample bias used in the measurements. This is also in agreement with previous STM measurements on fixed graphene resonator states, which did not experience effects from the probe tip potential (*14*).

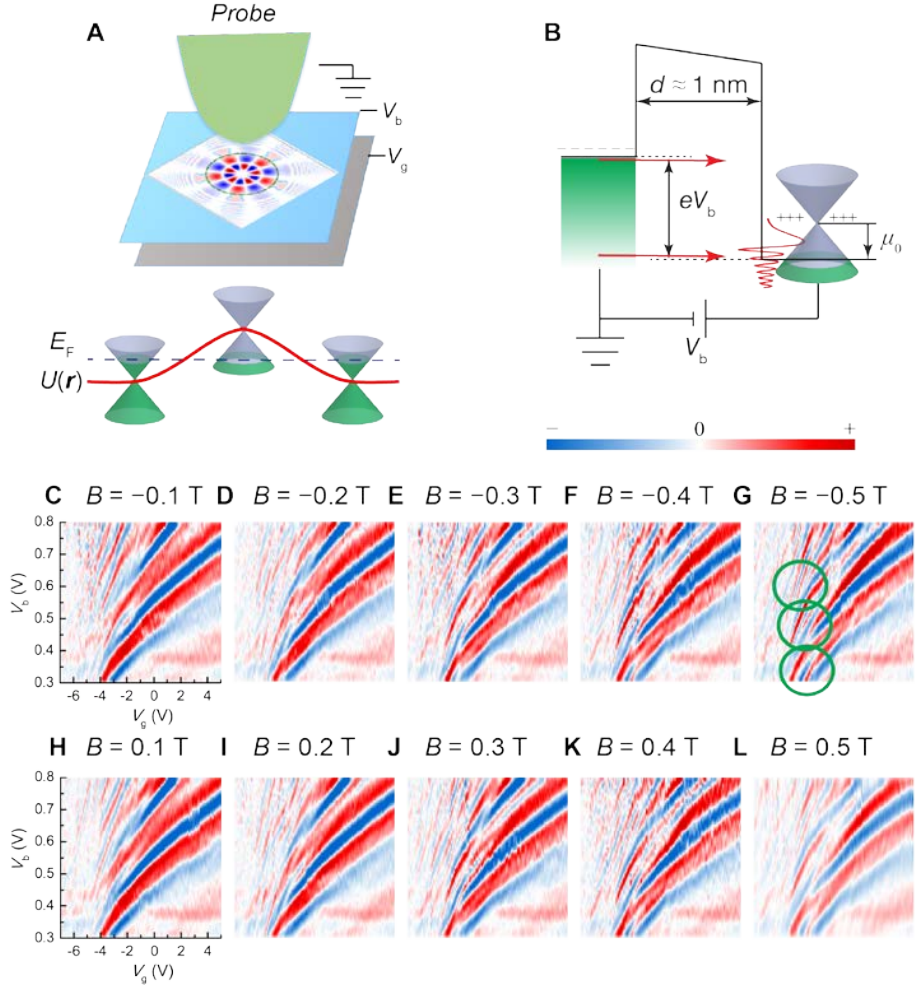

**Figure S3: Graphene resonator modes made by probe field-effect tip gating.** (**A**) Circular p-n junctions are created and probed simultaneously in the STM tunneling junction. The *p-n* junction ring is created by the electric field developed between the graphene sample bias $V_b$ and the grounded probe tip, which can invert the background density profile determined by the device back gate potential $V_g$, for sufficiently high sample biases. Klein scattering at oblique angles to the *p-n* junction boundary gives rise to confined whispering gallery modes (WGM) illustrated in the graphene plane (13). (**B**) Schematic tunnel diagram showing that the WGM modes appear in this portion of the gate map due to a 2nd tunneling channel aligned with the graphene Fermi level. The WGM are emptied one by one as the sample bias is increased, due to field-effect gating of the graphene, and can be seen as one traces up vertically in the gate maps. (**C**)-(**L**) Spectral maps consisting of $-d^2g/dV_b^2$ tunneling intensity as a function of back gate potential on the horizontal axis, $V_g = -7$ V to 5 V, and the sample potential on the vertical axis, $V_b = 0.3$ V to 0.8 V. Positive values are colored red, and negative blue. The first few WGM modes are seen as a collection of peaks (red) in the fans going from the lower left to upper right, and correspond to the WGM' modes in Ref. (13). The modes are seen to split into two peaks in various regions of the maps as a result of the Berry phase switch described in the main text. The green circles in (G) highlight the area where the splitting can be seen in the $B = -0.5$ T map, which is analyzed in detail in Fig. S4 and S5.

A. On/Off Berry phase switching of the traveling graphene resonator states

The measurements described in this section were made on the device described in reference (*13*). Figure S3A shows the geometry and Fig. S3,C-L shows the tunneling gate maps as a function of positive and negative magnetic fields. The resonator states are seen as a fan of peaks (red stripes) going from the bottom left to top right corner. These correspond to the whispering gallery modes (WGM) states, which result from tunneling from filled tip states at the graphene Fermi-level into empty WGM modes, which are empty due to the electric field gating from the probe, as described in Fig. S3B (*13*). Note that sample biases in excess of 300 mV are needed to observe the resonances. The new finding here is that these resonator modes split into two peaks in weak magnetic fields due to the turning on of the $\pi$ Berry phase, as highlighted by the green circles in Fig. S3G. The resonator states only split in a portion of the gate map because the potential strength is a function of the sample bias $V_b$ and back gate voltage $V_g$, and the critical field is dependent on the potential strength, as shown in Eq. (2) of the main text. This results in an almost linear boundary in the gate maps where the splitting will be observed on one side and not the other for a fixed magnetic field.

In Figure S4, we study the traveling dot resonator states in more detail, comparing $B = 0$ T (Fig. S4A) to $B = 0.5$ T (Fig. S4B) measurements. A close examination of the map at 0.5 T shows that the resonator modes are split by a large amount, on the order of half of their zero field spacing. We investigate the splitting of the modes by examining line cuts at fixed sample bias along the $V_g$ axis. Figure S5A shows a series of line cuts for the first $n = 0$ mode. A clear symmetric splitting of the mode is observed with magnetic field, reaching a value of ≈ 1 V. This gives a normalized splitting of approximately ½ taking the zero field mode spacing to be 2 V.

Figure S5B summarizes the mode splitting as a function of field for the $n = 0$ and $n = 1$ resonator modes. One can observe a low field transition, similar to Fig. 3D in the main text, but not with the same clarity due to the coarse field spacing of the measurements. Nevertheless, the measurements demonstrate that both graphene resonators made by impurity doping of the hBN insulator and probe field-effect gating show similar small critical magnetic fields, and energy jumps of approximately half the mode spacing, in agreement with theoretical predictions of a Berry phase switch.

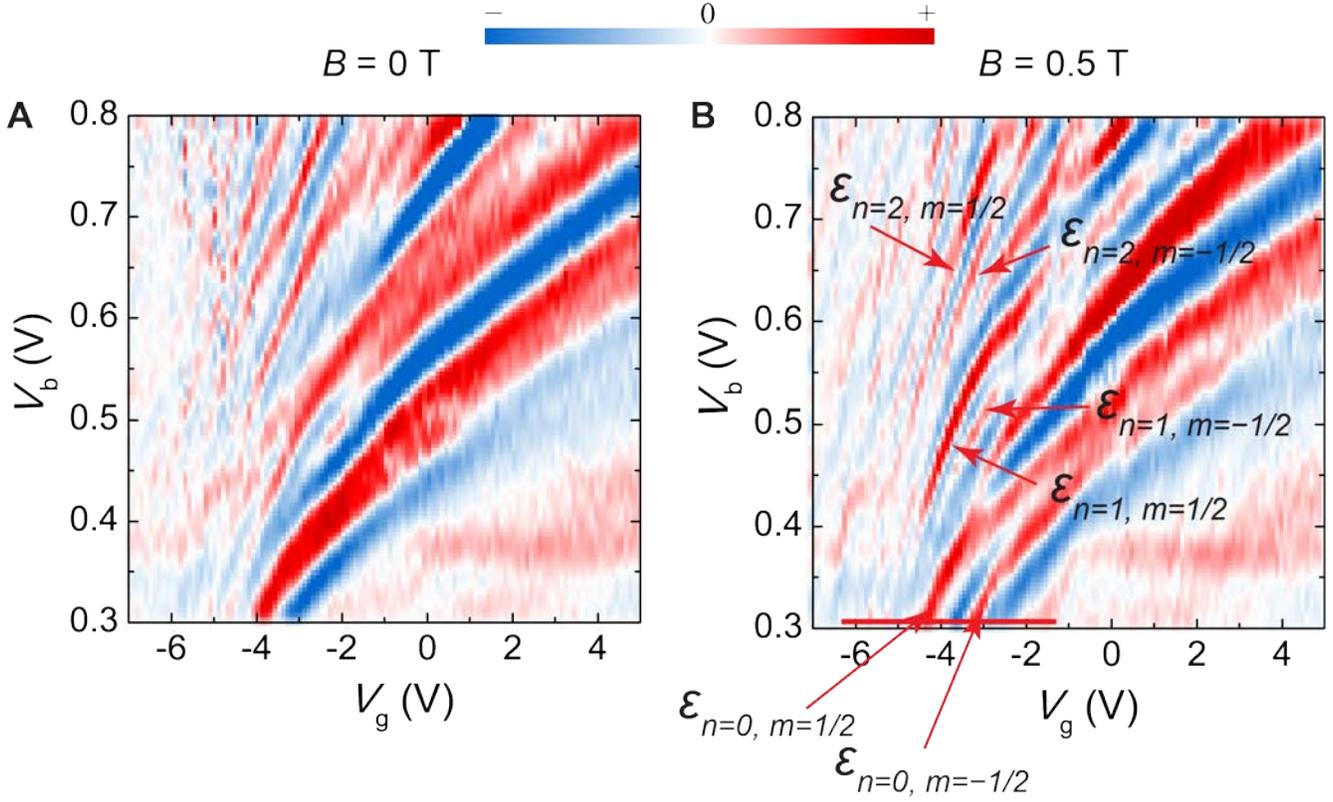

**Figure S4: Berry phase switching of graphene resonator modes made by field-effect tip gating.** Tunneling conductance gate maps at (**A**) $B = 0$ T and (*B*) $B = 0.5$ T illustrating the splitting of the graphene resonator states in weak magnetic fields. States corresponding to $n = 0$, 1, and 2 are seen to split in (B), as indicated by the red arrows, and analyzed further in Fig. S5. The maps are shown as $-d^2g/dV_b^2$ to remove the dispersive graphene background. Positive values are colored red, and negative blue. The critical field for Berry phase splitting depends on the potential strength (Eq. (2) in the main text). Hence, splitting is only observed in regions of the map where the potential is of sufficient strength forming a boundary in the left portion of the map in (B).

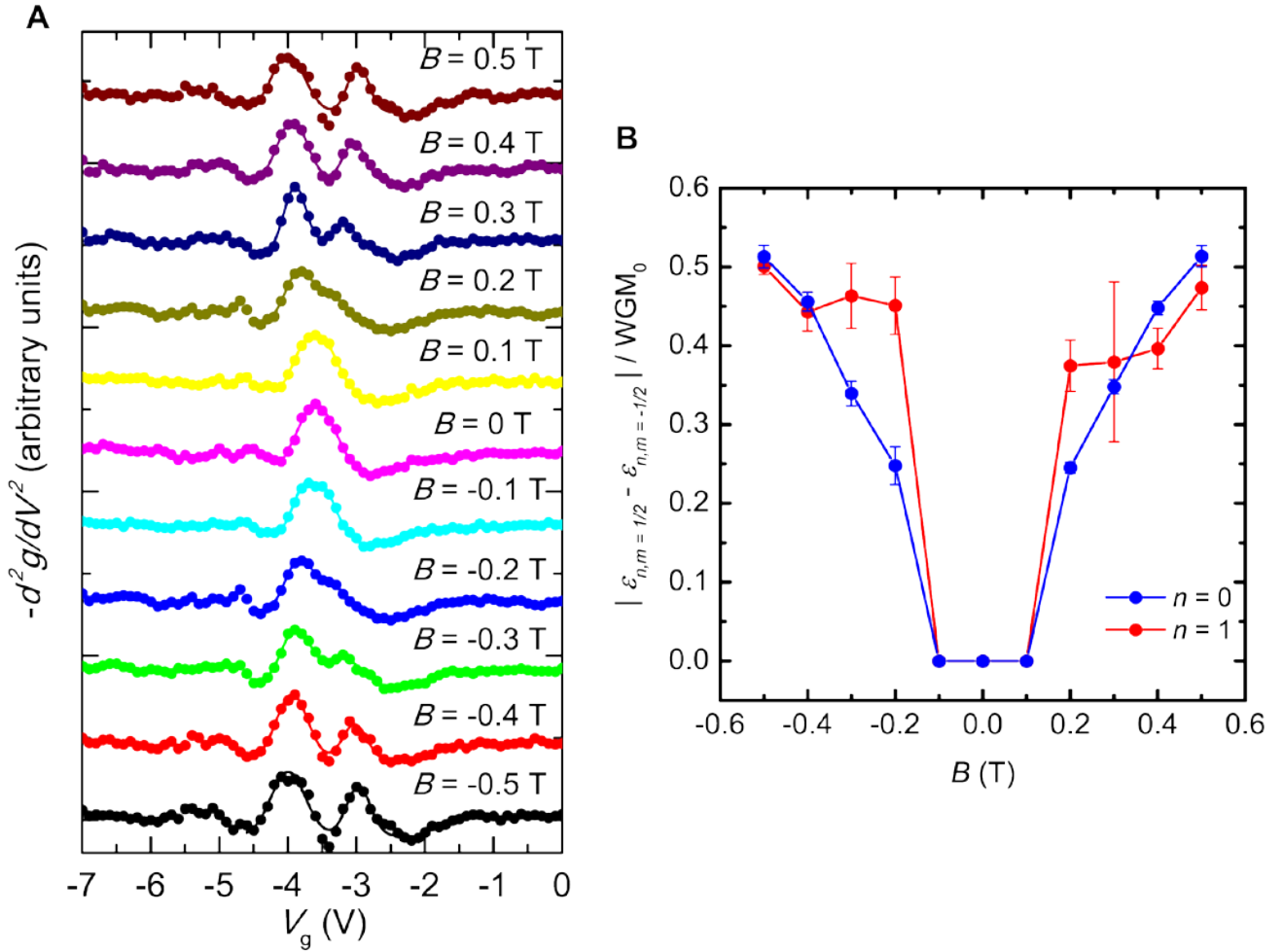

**Figure S5: Splitting of graphene resonator modes made by tip field-effect gating in weak magnetic fields.** (**A**) Line profiles (dots) at $V_b = 0.3$ V showing the splitting of the $n = 0$ mode vs. magnetic field obtained from the spectral maps in Fig. S4. The split peaks correspond to the $m = \pm 1/2$ states. The solid lines are fits to Gaussian functions used to extract the peak positions of the split resonator modes. (**B**) The difference in peak positions between the $m = \pm 1/2$ states, as obtained from (A), for the $n = 0$ and $n = 1$ modes normalized to the zero field mode spacing vs. magnetic field. The uncertainty reported represents one standard deviation and is determined by propagating the uncertainty resulting from a least square fit of Gaussian functions to determine the peak position of the resonator modes. A transition in the range of $B \approx \pm 0.2$ T is observed, and the energy difference reach magnitudes of $\approx 1/2$ of the zero field mode spacing, in agreement with the mode splitting observed in the fixed graphene resonators made by hBN impurity doping (Fig. 3D of the main text).

B. Discussion of the induced potential of the traveling graphene resonator

We calculate the induced potential in graphene to understand in detail how the induced potential varies with bias voltage and back gate voltage in the gate maps in Fig. S4. We use the Thomas Fermi approximation described in reference (*13*). Figure S6A shows the potential profiles (blue) relative to the Fermi level (orange lines) calculated using the model parameters in reference (*13*), for the region of the gate map shown in Fig. S4. A circular *p-n* junction is formed when the potential crosses the Fermi level, which only occurs for higher biases above 0.2 eV. In examining Fig. S6A, one can see that both the strength and size of the *p-n* junction vary with both $V_b$ and $V_g$. Figure S6B shows more clearly the potential profile along a line cut as a function of gate voltage at $V_b = 0.4\ eV$. One can see a significant variation in the width of the potential profile, which will affect the critical field via Eq. (2) of the main text.

Figure S7 shows the quadratic potential parameter $\kappa$, obtained from a fit of the curves in Fig. S6B over the range of ±50 nm. One can see a significant variation in $\kappa$ as a function of both gate voltage and bias voltage, with larger values (smaller *p-n* junction radius) tending to larger gate voltage and sample bias. This leads to a potential well which is dependent on position within the gate map. For this reason, we constructed the fixed *p-n* junctions to observe the Berry phase switching with measurements which are not influenced by the sample bias within the experimental range, and at fixed back gate voltage.

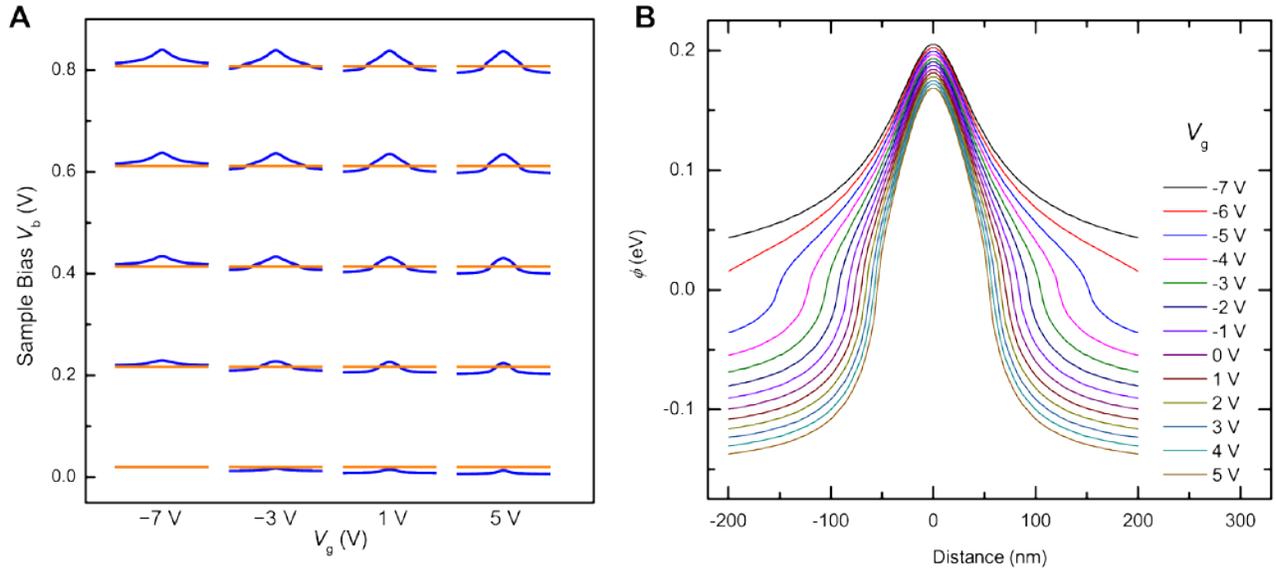

**Figure S6: The induced potential profile in the traveling graphene resonator**. (**A**) The induced potential (blue) is shown in the gate map range with the sample bias $V_b$, 0 V to 0.8 V, and back gate voltage $V_g$, -7 V to 5 V, to cover the range of the measurement in Fig. S4. Note that the *p-n* junctions are not defined until higher sample bias are reached when the blue curves cross the Fermi level shown by the orange lines. (**B**) The induced potential as a function of gate voltage at $V_b$=0.4 eV. The calculations use the parameters described in reference (13).

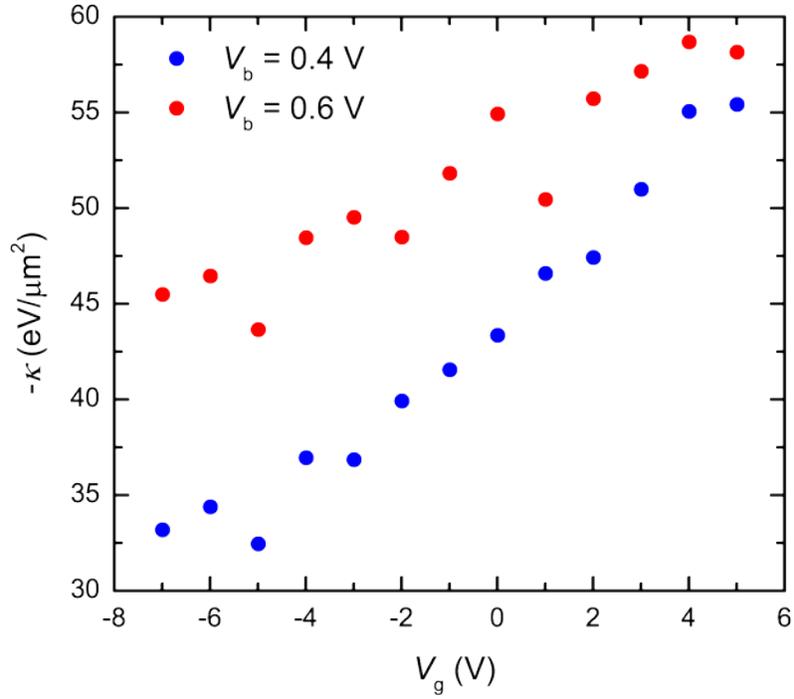

**Figure S7: Variation of the *p-n* junction potential radius in the traveling graphene resonator**. The quadratic potential parameter, $\kappa$, used to describe the confining potential is obtained from fitting the potential profiles as in Fig. S6 over the interval of ±50 nm, as a function of sample bias and back gate voltage. The parameter increases, which implies a smaller radius resonator, as the gate voltage and sample bias are increased.

## III. Circular Graphene Resonators Made with Impurity Doping of the hBN Insulator

Two methods were used to create to create circular *p-n* junction resonators in this study, using different graphene devices. In this section, we describe the device and measurements made by impurity doping the hexagonal boron nitride (hBN) insulating layer underneath the graphene. The device structure for the traveling *p-n* junction is described in detail in Ref. (13).

### A. Device fabrication and experimental set up

The graphene device used in the experiment in the main text was prepared by the method described in Ref. (*31*), where hBN is employed as a substrate. To fabricate graphene on hBN, single crystals of hBN are exfoliated onto 285 nm $SiO_2$/Si substrates where flakes with different thicknesses can be easily identified using an optical microscope. Selected exfoliated flakes were scanned by an AFM to check the surface roughness and to make sure they were free from tape residues and other contaminations. Graphene is exfoliated separately onto a stack consisting of polymethyl methacrylate (PMMA)/polyvinyl alcohol (PVA) /Si. The thickness of PMMA is tuned such that monolayer graphene (MLG) can be identified by an optical microscope. After floating the substrate on a deionized (DI) water bath, PVA (a water-soluble polymer) dissolves and the PMMA layer floats on top leaving the Si substrate at the bottom of the bath. Then, this membrane is picked up by a metal transfer slide and transferred on top of the target hBN flake using a micromanipulator fixed on an optical microscope. During transfer the substrate is heated to 120 ºC to facilitate the adhesion of PMMA to the substrate and also to remove the water absorbed on the surface of graphene and hBN before transfer. Figure S8A displays an image of a graphene/hBN stack prepared by this method. Raman spectroscopy on our device is shown in Fig. S8C. The sharpness and intensity of the graphene 2D peak relative to the G peak, $I_{2D}/I_G = 4$, confirms that the device is high-quality monolayer graphene (*32*). Note that hBN exhibits a characteristic Raman peak around 1366 cm$^{-1}$ and is very close to the graphene D peak (~1350 cm$^{-1}$) which is not present in our high-quality device. After transferring, Cr(1nm)/Pd(10nm)/Au(40nm) electrical contacts, including two fans of radial guidelines for STM navigation, were deposited employing a multi-step standard electron-beam lithography process (Fig. S8B). Finally, the device is annealed in a 5 % H$_2$/95 % Ar atmosphere at 350 ºC for several hours to remove the processing residues.

The sample is heated at 350 ºC in ultra-high vacuum for several hours before transferring it to the STM chamber. Tunneling measurements were performed in a custom ultra-high vacuum cryogenic STM system (*26*). The probe tip was optically aligned onto the graphene device before cooling to 4 K. Navigating from the landing spot to the graphene layer was accomplished with robotic walking algorithms (Fig. S8D). The $dI/dV_b$ spectroscopy measurements were carried out using a standard lock-in technique where a sinusoidal voltage amplitude of 2 mV RMS at 383Hz was added to the sample bias and the STM feedback loop disabled during measurement.

Figures S9, A and B show atomically-resolved topographic images of our graphene/hBN device and exemplify the cleanliness of the sample surface. The graphene-hBN moiré superlattice, which is not apparent in the topographic image in Fig.S9, A and B, can be most clearly seen in the fast Fourier transform (FFT) image in Fig. S9C (green circles). The magnitude and orientation of these superlattice peaks indicate a relative

misorientation of θ ≈ 29º between graphene and the hBN layer below, as confirmed by the simulated FFT in Fig. S9D.

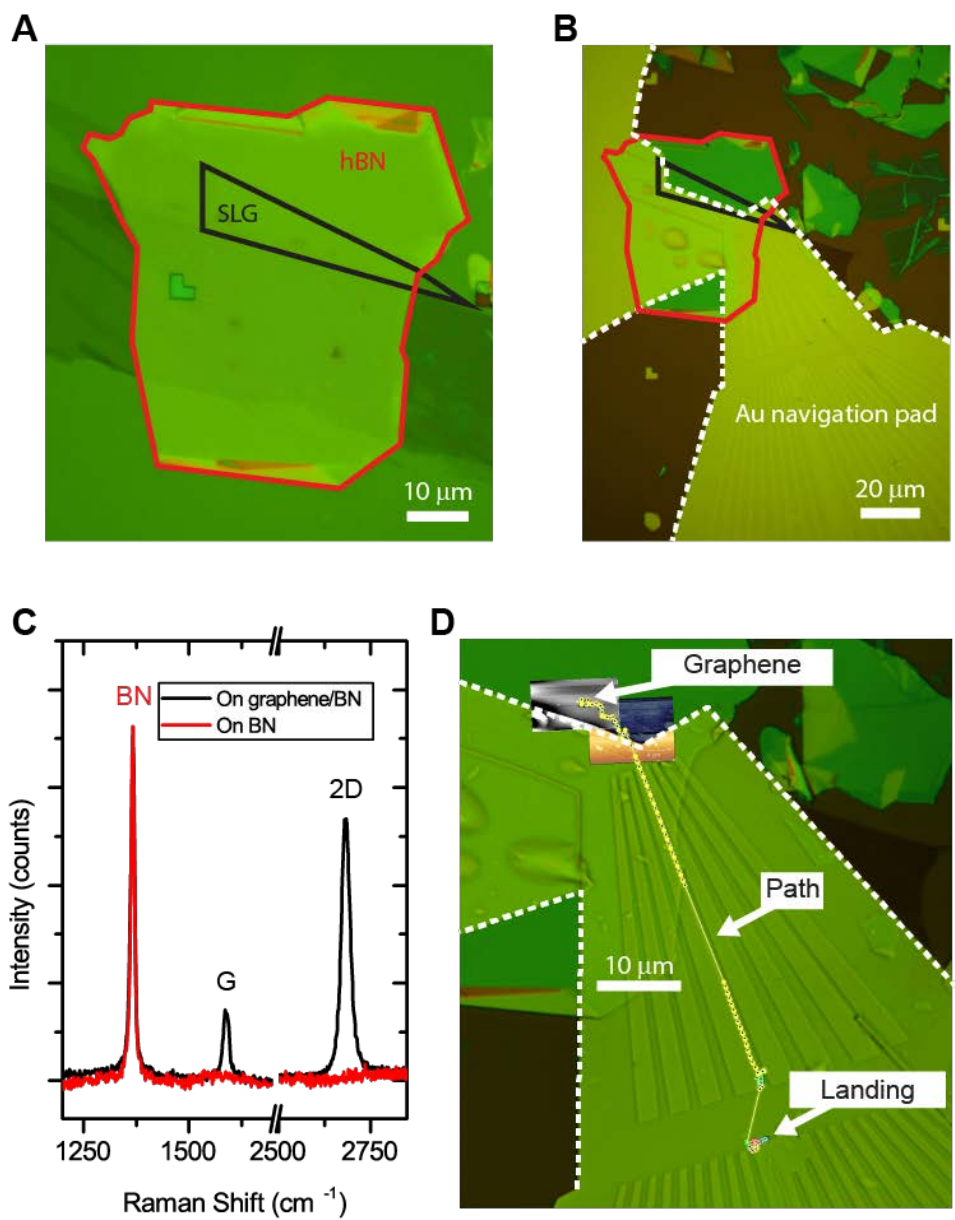

**Figure S8: Graphene device fabrication**. Optical image of a single-layer graphene (SLG) flake (black outline) on hBN (red outline) before (**A**) and after (**B**) deposition of gold contacts required for electrical contact and STM navigation to the SLG sample region. The dashed lines outline the boundary of metallic contact to graphene. (**C**) Raman spectrum recorded on graphene/BN (black) and on bare BN (red). The sharpness and intensity of the 2D peak is indicative of high-quality SLG. (**D**) Composite image comprising an optical micrograph of the navigation pad superimposed with AFM scans of the finished device and the path of the STM tip traveling from the landing spot to the graphene sample.

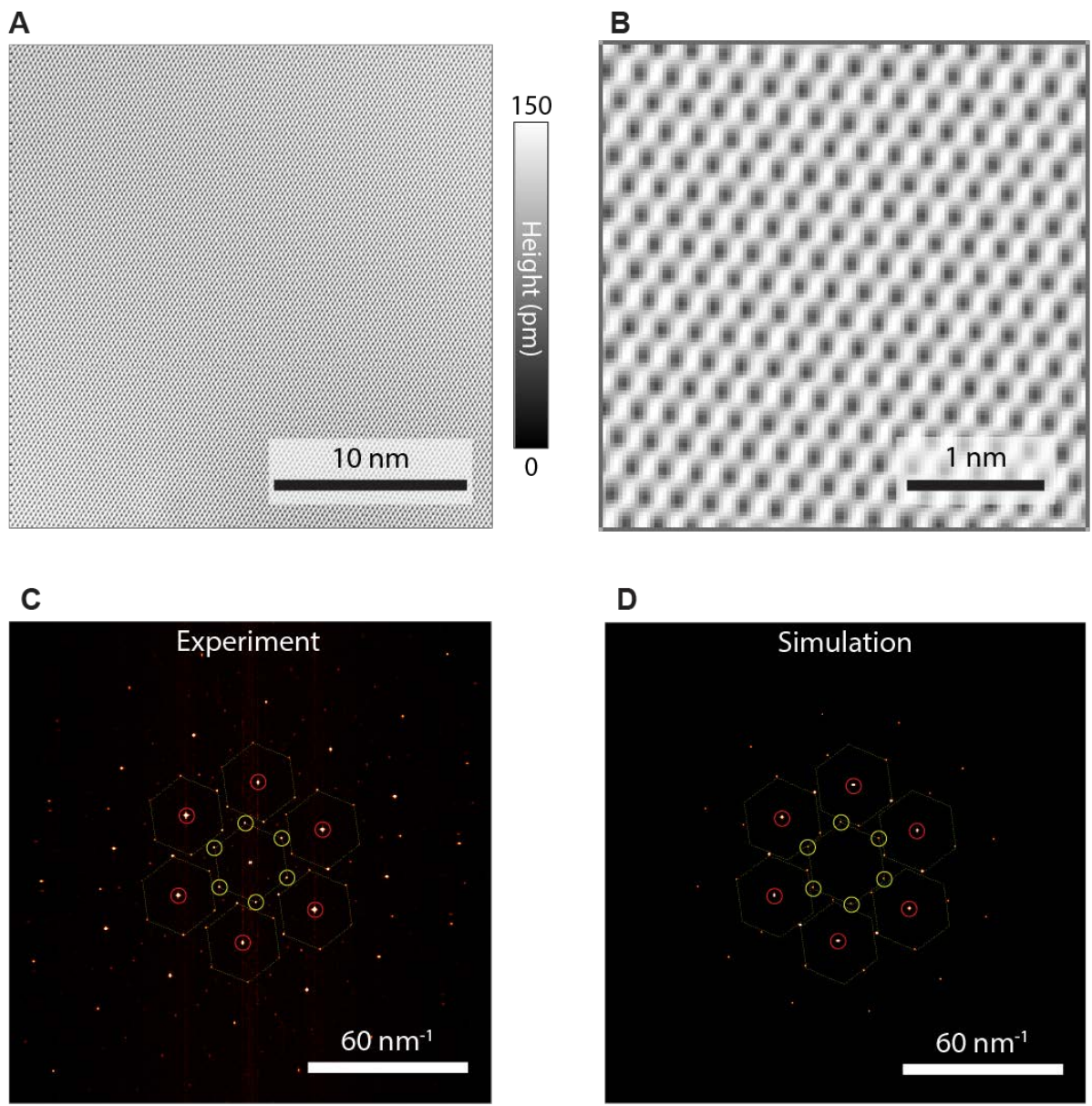

**Figure S9: Atomic-scale imaging of graphene and Moiré lattice determination**. (**A**) Large-scale atomically-resolved STM topograph of the graphene/hBN device displaying exceptional flatness and cleanliness of the sample surface ($V_b = 300$ mV, I = 300 pA, $V_g = 5$ V). (**B**) Enlarged view of a region of (A) highlighting the graphene honeycomb lattice. (**C**) Fast Fourier transform (FFT) of the experimental topograph in (A) displaying the graphene atomic Bragg peaks (red circles) as well as a short-wavelength Moiré superlattice (green circles) which also surrounds each Bragg peak (dotted green hexagons). The length and orientation of the moiré superlattice corresponds to an angular misorientation of $\theta \approx 29°$ between the graphene and the underlying hBN layer. (**D**) FFT of a simulated STM topograph of graphene and hBN misoriented at $\theta = 29.31°$ to compare with (C), which reproduces the primary features of the experimental topograph.

A. Creating and characterizing fixed circular *p-n* junctions

Here we apply a patterning technique developed in Ref. (*28*) to create a fixed circular *p-n* junction in a graphene/hBN device on top of a $SiO_2$/Si substrate. Figure S10 (top panel) displays a schematic of the sample and measurement geometry. Figure S10 (lower panel) describes the method used to create the *p-n* junctions. First, (*i*) neutral hBN impurities are (*ii*) subjected to an external electric field, $E_g$, produced by a back gate voltage of $V_g = 30$ V. Next, (*iii*) the STM tip is retracted ≈2 nm above the graphene surface and the sample bias relative to the grounded tip is ramped to 5 V and held for a time $t \approx 60$ s. This voltage pulse ionizes the impurities residing in hBN underneath the tip (*28*). The ionized impurities and charges redistribute themselves to create an internal electric field, $E_d$, which tends to cancel the external back gate electric field. Finally, (*iv*) when the external gate voltage is removed (or lowered), the internal impurity field inside the hBN acts as a local negative embedded gate which induces positive charges (holes) in the graphene layer above. The global back gate voltage $V_g$ can then be tuned such that the graphene is overall *n*-doped except for the circular *p*-doped region where the voltage pulse was performed. (Note that an *n*-doped circular dot can be created the same way by applying a negative back gate voltage during the voltage pulse, as demonstrated in Refs. (*14*, *28*)).

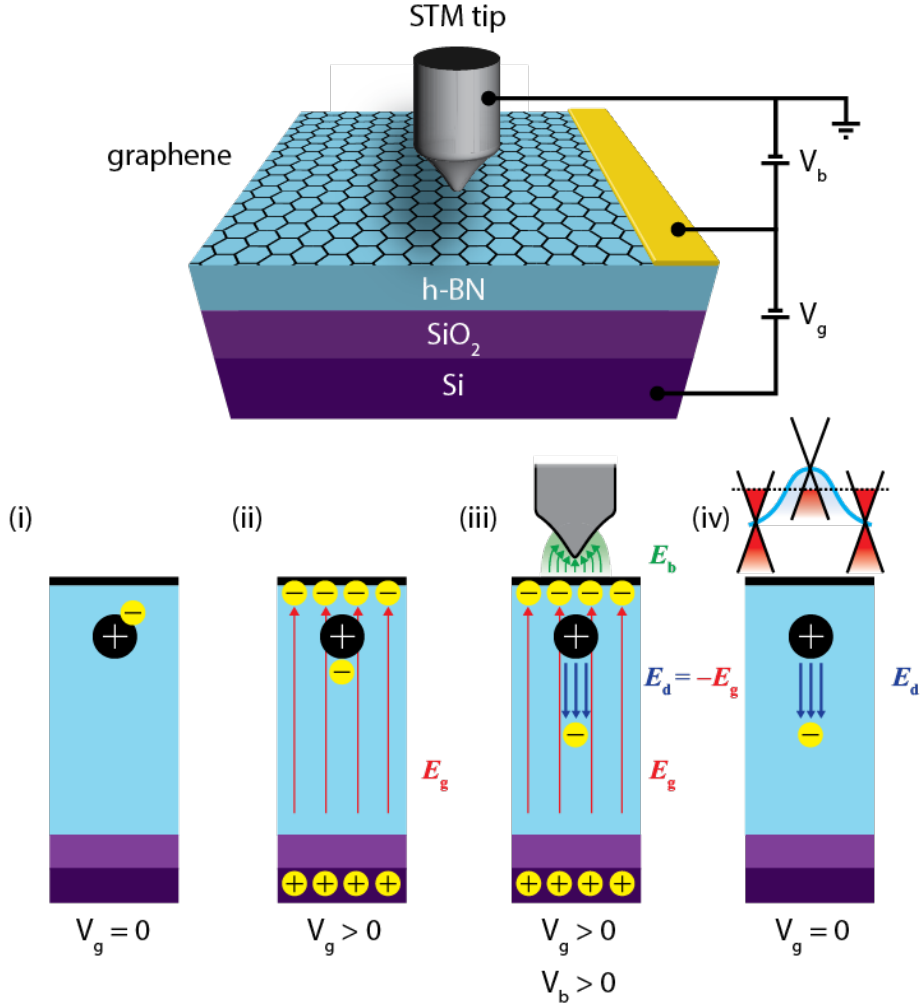

**Figure S10: Creating fixed graphene quantum dots with charged impurity doping of hBN**. (Top) Schematic of the graphene device and measurement setup. The device consists of graphene on a substrate consisting of hBN (blue)/SiO$_2$ (light purple)/Si (dark purple). (Bottom) The creation of the quantum dot follows the method in Refs. (14, 28) and proceeds as follows: (*i*) Neutral impurities reside within the hBN. (*ii*) A positive backgate voltage, $V_g$ = 30 V, is applied between silicon and graphene. This electric backgate field, $E_g$ (red field lines), draws negative charge into graphene, producing n-doping. (*iii*) A voltage pulse of $V_b$ = 5 V is applied between the STM tip and graphene for a time t = 60 s. The voltage pulse ionizes hBN impurities directly below the STM tip. The ionized impurities and released charge rearrange to create an opposite field, $E_d$ (blue field lines), to screen the external gate field. (*iv*) When the external gate is removed (or lowered), the charged impurities remain and act as an embedded gate, locally *p*-doping the graphene.

B. Sharpness of the On/Off Critical Field Threshold

In Figure S11, we examine the sharpness of the On/Off switching of the resonator states at the critical magnetic field. Figure S11A shows the magnified view of the $n = 4$ resonance conductance map from Fig. 3C. Overlaid on the map in yellow are the traces of $-d^2g/dV_b^2$ at $B = -0.06$ T and $B = -0.252$ T, displaying a single peak and a double peaked spectrum, respectively. In Figure S11B, we plot the spectra located in the oval region of Fig. S11A, to examine in detail the Berry phase switching at negative magnetic fields, covering the field range of $B = -0.092$ T to $-0.132$ T, in steps of 0.004 T. We see that the spectra changes from a single peak at lower negative fields, corresponding to the Berry phase of zero, to a double peaked spectrum, where the Berry phase equals $\pi$, when the field changes from $B = -0.104$ T to $-0.108$ T. A discernable $2^{nd}$ peak is not present at $B = -0.096$ T, but pops out of the background at $B = -0.108$ T. This corresponds to a magnetic field change of 0.012 T, and shows that the On/Off Berry phase transition can be modulated with an alternating magnetic field on the order of 10 mT.

C. Off-Center Spectra

The differential conductance spectra measured off center show the switching of higher $m$ states at higher critical magnetic fields. In Fig. S12 we include more spectra measured off center, which show similar critical field switching as in Fig. 4A, but also show how the states can be viewed as connected going downward or upward in energy depending on fine details of the transition regions (see discussion in main text).

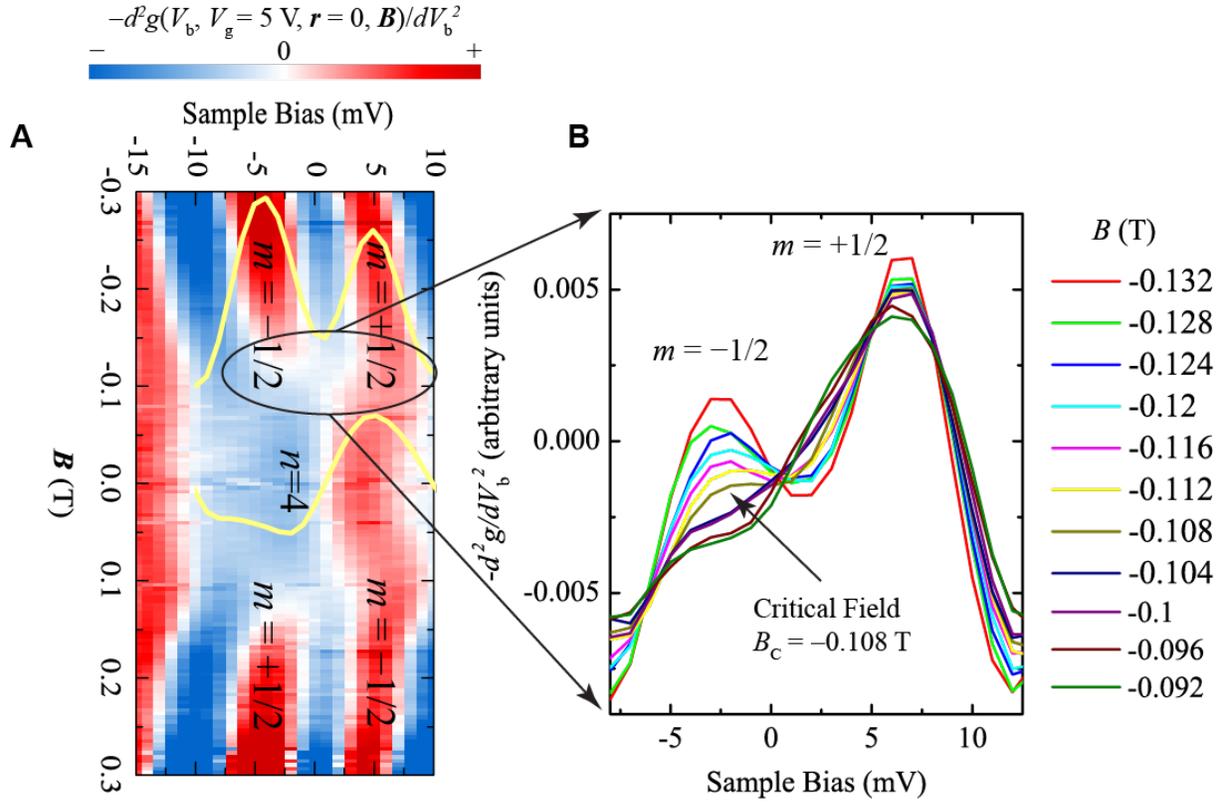

**Figure S11: Sharpness of the On/Off Critical Field Threshold**. (**A**) Magnified view of the $n = 4$ resonance from Fig. 3A in the main text, showing the asymmetric splitting of the $m = \pm 1/2$ modes with magnetic field. Traces of $-d^2g/dV_b^2$ at $B=-0.06$ T and $B=-0.252$ T, displaying a single peak, and a double peaked spectrum, respectively, are shown overlaid on the map (yellow lines). (**B**) A series of differential conductance spectra vs. sample bias for different magnetic fields covering the negative critical field transition shown by the oval in (A). At the critical field, $B_C=-0.108$ T, a second peak corresponding to the $m=-1/2$ resonance suddenly appears above the background signal, signifying the On/Off Berry phase switching transition is very sharp.

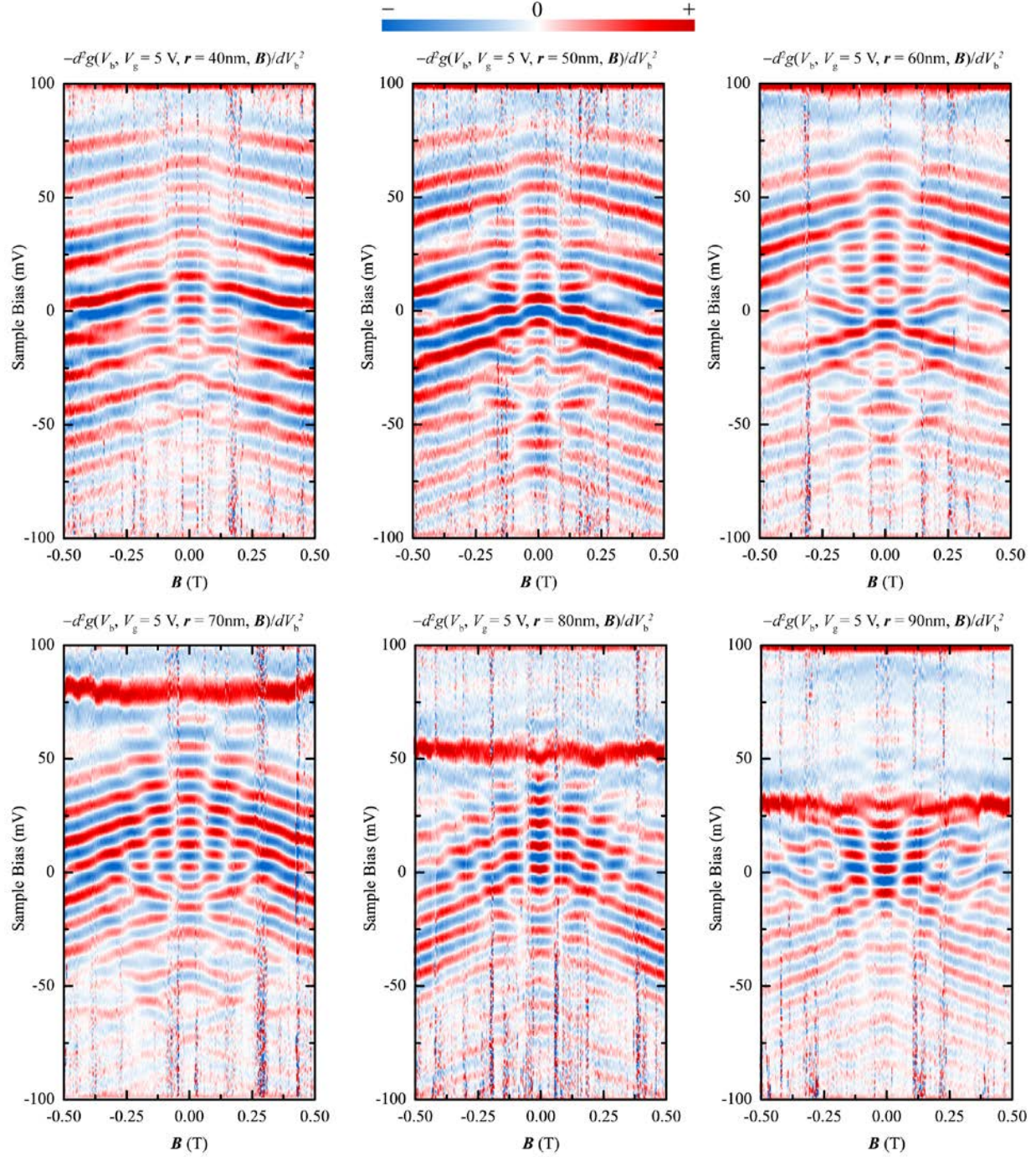

**Figure S12: Off-center spectral maps.** Differential conductance maps vs. magnetic field measured at distances of 40 nm to 90 nm from the center of the graphene resonator. States with higher *m* modes have more weight at positions farther away from the center of the *p-n* junction ring, where multiple transitions are observed with increasing magnetic field consistent with predictions from Eq. (2) in the main text. Each map consists of a series of 321 spectra plotted vertically as a function of magnetic field. The maps are shown in the 2nd derivative (arbitrary values). Positive values are colored red, and negative blue.

## IV. Full Reference List

**Movie S1: Evolution of graphene resonator orbits through the critical field**. (**Left**) Calculated graphene electron orbits in a quadratic potential with angular momentum states $m = -1/2$ (blue) and $m = +1/2$ (red). The initial radial position (at the inner turning point) is chosen so that the radial action $J_r \equiv \oint dr\, p_r(r) = h/2$ for all frames, and the particle orbits for 1 ps. Parameters are $v_F = 10^6$ m/s, $\kappa = 10\ eV/\mu m^2$ [see Section 1B]. (**Right**) Above, the kinematic momentum **Π**, as evaluated along $C_r$ described in the main text, for the trajectories shown on the left. The momentum contours are plotted on the Dirac cone (brown) using the kinetic energy $T(\mathbf{\Pi}) = v_F|\mathbf{\Pi}|$. The **Π**-values plotted are the dynamically calculated radial and azimuthal momentum during the orbit. Below, a plot of Berry phase vs. magnetic field for the two orbits.

**References and Notes**

1. M. V. Berry, Quantal Phase Factors Accompanying Adiabatic Changes. *Proc. R. Soc. Lond. Math. Phys. Eng. Sci.* **392**, 45–57 (1984).

2. M. Berry, Anticipations of the Geometric Phase. *Phys. Today* **43**, 34–40 (1990). doi:10.1063/1.881219

3. F. Wilczek, A. Shapere, *Geometric Phases in Physics* (World Scientific, 1989; www.worldscientific.com/worldscibooks/10.1142/0613), vol. 5 of *Advanced Series in Mathematical Physics*.

4. J. W. Zwanziger, M. Koenig, A. Pines, Berry's Phase. *Annu. Rev. Phys. Chem.* **41**, 601–646 (1990). doi:10.1146/annurev.pc.41.100190.003125

5. D. Xiao, M.-C. Chang, Q. Niu, Berry phase effects on electronic properties. *Rev. Mod. Phys.* **82**, 1959–2007 (2010). doi:10.1103/RevModPhys.82.1959

6. T. Bitter, D. Dubbers, Manifestation of Berry's topological phase in neutron spin rotation. *Phys. Rev. Lett.* **59**, 251–254 (1987). doi:10.1103/PhysRevLett.59.251 Medline

7. R. Tycko, Adiabatic rotational splittings and Berry's phase in nuclear quadrupole resonance. *Phys. Rev. Lett.* **58**, 2281–2284 (1987). doi:10.1103/PhysRevLett.58.2281 Medline

8. R. Y. Chiao, Y.-S. Wu, Manifestations of Berry's topological phase for the photon. *Phys. Rev. Lett.* **57**, 933–936 (1986). doi:10.1103/PhysRevLett.57.933 Medline

9. A. Tomita, R. Y. Chiao, Observation of Berry's topological phase by use of an optical fiber. *Phys. Rev. Lett.* **57**, 937–940 (1986). doi:10.1103/PhysRevLett.57.937 Medline

10. K. S. Novoselov, A. K. Geim, S. V. Morozov, D. Jiang, M. I. Katsnelson, I. V. Grigorieva, S. V. Dubonos, A. A. Firsov, Two-dimensional gas of massless Dirac fermions in graphene. *Nature* **438**, 197–200 (2005). doi:10.1038/nature04233 Medline

11. Y. Zhang, Y.-W. Tan, H. L. Stormer, P. Kim, Experimental observation of the quantum Hall effect and Berry's phase in graphene. *Nature* **438**, 201–204 (2005). doi:10.1038/nature04235 Medline

12. D. L. Miller, K. D. Kubista, G. M. Rutter, M. Ruan, W. A. de Heer, P. N. First, J. A. Stroscio, Observing the quantization of zero mass carriers in graphene. *Science* **324**, 924–927 (2009). doi:10.1126/science.1171810 Medline

13. Y. Zhao, J. Wyrick, F. D. Natterer, J. F. Rodriguez-Nieva, C. Lewandowski, K. Watanabe, T. Taniguchi, L. S. Levitov, N. B. Zhitenev, J. A. Stroscio, Physics. Creating and probing electron whispering-gallery modes in graphene. *Science* **348**, 672–675 (2015). doi:10.1126/science.aaa7469 Medline

14. J. Lee, D. Wong, J. Velasco Jr., J. F. Rodriguez-Nieva, S. Kahn, H.-Z. Tsai, T. Taniguchi, K. Watanabe, A. Zettl, F. Wang, L. S. Levitov, M. F. Crommie, Imaging electrostatically confined Dirac fermions in graphene quantum dots. *Nat. Phys.* **12**, 1032–1036 (2016). doi:10.1038/nphys3805

15. C. Gutiérrez, L. Brown, C.-J. Kim, J. Park, A. N. Pasupathy, Klein tunnelling and electron trapping in nanometre-scale graphene quantum dots. *Nat. Phys.* **12**, 1069–1075 (2016). doi:10.1038/nphys3806

16. J. F. Rodriguez-Nieva, L. S. Levitov, Berry phase jumps and giant nonreciprocity in Dirac quantum dots. *Phys. Rev. B* **94**, 235406 (2016). doi:10.1103/PhysRevB.94.235406

17. A. V. Shytov, M. S. Rudner, L. S. Levitov, Klein backscattering and Fabry-Pérot interference in graphene heterojunctions. *Phys. Rev. Lett.* **101**, 156804 (2008). doi:10.1103/PhysRevLett.101.156804 Medline

18. A. F. Young, P. Kim, Quantum interference and Klein tunnelling in graphene heterojunctions. *Nat. Phys.* **5**, 222–226 (2009). doi:10.1038/nphys1198

19. V. V. Cheianov, V. Fal'ko, B. L. Altshuler, The focusing of electron flow and a Veselago lens in graphene p-n junctions. *Science* **315**, 1252–1255 (2007). doi:10.1126/science.1138020 Medline

20. A. De Martino, L. Dell'Anna, R. Egger, Magnetic confinement of massless Dirac fermions in graphene. *Phys. Rev. Lett.* **98**, 066802 (2007). doi:10.1103/PhysRevLett.98.066802 Medline

21. P. E. Allain, J. N. Fuchs, Klein tunneling in graphene: Optics with massless electrons. *Eur. Phys. J. B* **83**, 301–317 (2011). doi:10.1140/epjb/e2011-20351-3

22. L. C. Campos, A. F. Young, K. Surakitbovorn, K. Watanabe, T. Taniguchi, P. Jarillo-Herrero, Quantum and classical confinement of resonant states in a trilayer graphene Fabry-Pérot interferometer. *Nat. Commun.* **3**, 1239 (2012). doi:10.1038/ncomms2243 Medline

23. A. Varlet, M.-H. Liu, V. Krueckl, D. Bischoff, P. Simonet, K. Watanabe, T. Taniguchi, K. Richter, K. Ensslin, T. Ihn, Fabry-Pérot interference in gapped bilayer graphene with broken anti-Klein tunneling. *Phys. Rev. Lett.* **113**, 116601 (2014). doi:10.1103/PhysRevLett.113.116601 Medline

24. J.-S. Wu, M. M. Fogler, Scattering of two-dimensional massless Dirac electrons by a circular potential barrier. *Phys. Rev. B* **90**, 235402 (2014). doi:10.1103/PhysRevB.90.235402

25. N. A. Garg, S. Ghosh, M. Sharma, Scattering of massless Dirac fermions in circular p-n junctions with and without magnetic field. *J. Phys. Condens. Matter* **26**, 155301 (2014). doi:10.1088/0953-8984/26/15/155301 Medline

26. R. J. Celotta, S. B. Balakirsky, A. P. Fein, F. M. Hess, G. M. Rutter, J. A. Stroscio, Invited Article: Autonomous assembly of atomically perfect nanostructures using a scanning tunneling microscope. *Rev. Sci. Instrum.* **85**, 121301 (2014). doi:10.1063/1.4902536 Medline

27. See additional supplementary text and data.

28. J. Velasco Jr., L. Ju, D. Wong, S. Kahn, J. Lee, H.-Z. Tsai, C. Germany, S. Wickenburg, J. Lu, T. Taniguchi, K. Watanabe, A. Zettl, F. Wang, M. F. Crommie, Nanoscale Control of Rewriteable Doping Patterns in Pristine Graphene/Boron Nitride Heterostructures. *Nano Lett.* **16**, 1620–1625 (2016). doi:10.1021/acs.nanolett.5b04441 Medline

29. A. Einstein, On the quantum theorem of Sommerfeld and Epstein. *Deutshe Phys. Ges.* **19**, 82–92 (1917).

30. A. D. Stone, Einstein's unknown insight and the problem of quantizing chaos. *Phys. Today* **58**, 37–43 (2005). doi:10.1063/1.2062917

31. C. R. Dean, A. F. Young, I. Meric, C. Lee, L. Wang, S. Sorgenfrei, K. Watanabe, T. Taniguchi, P. Kim, K. L. Shepard, J. Hone, Boron nitride substrates for high-quality graphene electronics. *Nat. Nanotechnol.* **5**, 722–726 (2010). doi:10.1038/nnano.2010.172 Medline

32. A. C. Ferrari, J. C. Meyer, V. Scardaci, C. Casiraghi, M. Lazzeri, F. Mauri, S. Piscanec, D. Jiang, K. S. Novoselov, S. Roth, A. K. Geim, Raman spectrum of graphene and graphene layers. *Phys. Rev. Lett.* **97**, 187401 (2006). doi:10.1103/PhysRevLett.97.187401 Medline

31